\documentclass[iop]{emulateapj}

\usepackage{graphicx}
\usepackage{hyperref}
\usepackage{amsmath}
\usepackage{color}

\shorttitle{Planetary Temperature Effects on Circumplanetary Disks}
\shortauthors{Szul\'agyi}

\begin{document}

\title{Effects of the Planetary Temperature on the Circumplanetary Disk and on the Gap}


\author{J. Szul\'agyi$^{1}$}
\affil{$^{1}$ETH Z\"urich, Institute for Astronomy, Wolfgang-Pauli-Strasse 27, Z\"urich, CH-8093, Switzerland}

\email{judits@ethz.ch}

\begin{abstract}
Circumplanetary disks regulate the late accretion to the giant planet and serve as the birthplace for satellites. Understanding their characteristics via simulations also helps to prepare for their observations. Here we study disks around 1, 3, 5, 10 $\mathrm{M_{Jup}}$ planets with three dimensional, global radiative hydrodynamic simulations with sub-planet peak resolution, and various planetary temperatures. We found that as the 1 $\mathrm{M_{Jup}}$ planet radiates away its formation heat, the circumplanetary envelope transitions to a disk between $T_p = 6000$K and 4000K.  In the case of 3-10 $\mathrm{M_{Jup}}$ planets a disk always forms. The temperature profile of the circumplanetary disks is very steep, the inner 1/6th is over the silicate condensation temperature and the entire disk is above water freezing point, making satellite formation impossible in this early stage ($<$1 Myr). Satellites might form much later and first in the outer parts of the disk migrating inwards later on. Our disk masses are 1, 7, 20, 40$\times 10^{-3}\mathrm{M_{Jup}}$ for the 1, 3, 5, 10 $\mathrm{M_{Jup}}$ gas giants respectively, and we provide an empirical formula to estimate the subdisk masses based on the planet- and circumstellar disk mass. Our finding is that the cooler the planet, the lower the temperature of the subdisk, the higher the vertical influx velocities, and the planetary gap is both deeper and wider. We also show that the gaps in 2D and 3D are different. The subdisk eccentricity increases with $\mathrm{M_p}$ and violently interacts with the circumstellar disk, making satellite-formation less likely, if $\mathrm{M_p} \gtrsim 5 \mathrm{M_{Jup}}$.
\end{abstract}


\section{Introduction} \label{sec:intro}

During the final planet formation phase, giant planets are surrounded by their own disk, the circumplanetary disks. We also call these subdisks, referring to the fact that they exist {\it within} the circumstellar disk. Their importance lies in the fact that they are regulating the late gas accretion to the planet, and they are the birthplace for satellites to form. Therefore both for planet formation and satellite formation perspective we need to understand their characteristics better. With no clear observational evidence of such subdisks until today, we have to rely on hydrodynamic simulations to unveil their properties. 

We are just entering an era when the detection of the circumplanetary disk is possible. Extended thermal emission around embedded, young, forming planets were already detected with direct imaging observations in a number of sources \citep[e.g.][]{KI12,Quanz15,Brittain14,Reggiani14,Sallum15}. These extended thermal emissions suggest that these young planets are surrounded by hot gas and dust, likely in the form of circumplanetary disks or -envelopes. Recently, \citet{Sallum15} detected $\mathrm{H}\alpha$ emission from one of the planetary candidates around LkCa15, which accretion tracer can either arise from the planet or its circumplanetary disk if it exists \citep{SzM16}. Furthermore, in the near future it will be also possible to detect the dust emission of the subdisk with the Atacama Large Millimeter Array (Pineda et al. in prep., Szulagyi et al. in prep.) or the kinematic imprints of this disk \citep{Perez15}. In order to prepare for future observations, we need models to predict the basic characteristics -- such as mass, temperature, kinematic properties -- of the circumplanetary disks. This is also essential for choosing the right method to detect these subdisks, e.g. via direct imaging, dust emission observations, or line broadening of certain spectral lines.

So far hydrodynamic studies of subdisks reported quite different characteristics. Regarding their mass, disk instability simulations of \citet{SB13} and \citet{Galvagni12} found very massive, $\sim 0.25-1.0 \mathrm{M_{planet}}$ circumplanetary disks. On the other hand, all core-accretion works measured masses around $10^{-3}-10^{-4} \mathrm{M_{planet}}$ \citep{Gressel13,Szulagyi14,Szulagyi16}. \citet{SzMayer16} showed that this discrepancy originates from the significantly different circumstellar disk masses between the two formation scenarios. Disk instability requires a very massive circumstellar disk ($>0.1 \mathrm{M_{Solar}}$), while the core-accretion simulations usually work with at least ten times lighter protoplanetary disks. According to \citet{SzMayer16}, the circumplanetary disks around core-accretion formed planets can be almost as massive as the disk instability formed ones, if the circumstellar disks have a similar mass.

Temperature-wise, core-accretion formed subdisks are an order of magnitude hotter \citep{AB09b,DA03,Gressel13,Szulagyi16} than their gravitational instability formed counterparts \citep{SB13,Galvagni12,SzMayer16}. For a Jupiter-mass planet, \citet{AB09b} found 4500 K peak temperature in the circumplanetary disk if they assumed realistic  planetary radius. When they defined the surface of the planet further out, at 0.02 $\mathrm{R_{Hill}}$, the inner circumplanetary disk temperature was only 1600 K. \citet{Szulagyi16} found over 13000\,K peak temperature for a Jupiter-mass planet with a sub-planet resolution of 80\% Jupiter-diameter. This highlights that the resolution, i.e. how close one gets to the planetary surface, how peaky is the gravitational potential well of the planet, matters a lot for the temperature in the planet vicinity. Other factors, such as viscosity \citep{DA03} and opacity \citep{PN05,DA14} are also playing a role in affecting the heat budget of the subdisk. \citet{DA03} showed that larger viscosity means higher temperatures in the circumplanetary disk, due to larger shear-stress by viscous forces. The magneto-hydrodynamic simulations of \citet{Gressel13} studied a bit smaller mass planets, growing them from 100 $\mathrm{M_{Earth}}$ to 150 $\mathrm{M_{Earth}}$ and already at these low-mass cores resulted in peak temperatures over 1500-2000 K in the subdisk. Similarly, the characteristic temperatures in the circumplanetary disk were 1000-2000\,K in the work of \citet{PN05} at various dust-to-gas ratio (1\% and lower). The peak temperatures are however not representing the bulk temperature of the subdisk, the temperature-profile sharply decreases with the distance from the planet \citep{AB09a,AB09b,DA03,PN05,DA14,Szulagyi16}.

With larger planetary mass ($\gtrsim 3 \mathrm{M_{Jup}}$), the gaps become more eccentric, therefore the circumplanetary disk is also more elongated \citep[e.g.][]{Lubow99,Kley99,LDA06,KD06}. As \citet{KD06} showed, in the $\gtrsim 5 \mathrm{M_{Jup}}$ cases the circumstellar disk can regularly engulf the subdisk, creating accretional bursts to the planet. Circumplanetary disks around these massive planets are particular interesting for observational point of view, because they likely to be more massive and more luminous than their counterparts around Jupiter-mass planets or below \citep[e.g.][]{Zhu15,SzM16}.

The temperature of circumplanetary disks can infer when and where satellite formation can occur inside these disks. First of all, the temperature should be below the dust (silicate) sublimation point ($\sim 1500$\,K) in order to have any hope to form satellites. Second, from the water ice content of Galilean satellites it is suggested that they had to form in a subdisk below water's freezing point \citep[e.g.][]{CW02}. This, at first glance seems to disprove the core accretion formed circumplanetary disk simulations with significantly higher temperatures inside the subdisk. However, as the planet radiates away its formation heat and the circumstellar disk gas content is dissipating, presumably the gas-giant's vicinity also cools off, eventually reaching low enough temperature for satellite formation to occur. The timescale of this is still unknown \citep{CW02,CW06,Estrada09,ME03a,ME03b} and certainly depends on number of factors, such as semi-major axis, opacity of the gas and dust, the temperature and cooling rate of the planet. Motivated by this and the observational efforts of the circumplanetary disks, in this paper we study how the planetary temperature affects the circumplanetary gas, namely its temperature, mass, and kinematic properties. With a suite of sub-planet resolution simulations with decreasing planetary temperatures for a 1 Jupiter-mass planet we study how these disk properties change. In the different calculations, we set the maximal planet temperature to 10000 K, 8000 K, 6000 K, 4000 K, 2000 K, 1000 K according to planet interior \& evolution studies (Mordasini et al. in prep., \citealt{Guillot95}). This temperature sequence can also be understood as an evolutionary ordering as the planet cools off, the lowest temperatures representing ages around 1-2 Myrs according to Mordasini et al. in prep. and \citet{Guillot95}. We also carried out simulations for 3, 5, 10 $\mathrm{M_{Jup}}$ planets with and without temperature cap. Due to the computationally very expensive simulations, we could not follow such a temperature sequence as in the case of the 1 $\mathrm{M_{Jup}}$ planet, but we simulated 2 cases: non fixed temperature ($>$12000 K) and fixed temperature with 4000\,K. With all the simulations presented in this paper and in \citet{Szulagyi16}, \citet{SzMayer16} we also created an empirical formula to estimate the circumplanetary disk mass for observations, based on the protoplanetary disk mass and the planet mass. The found subdisk characteristics, furthermore, can provide predictions for observational efforts to detect these disks, as well as to help understand when and where satellites could form inside them.

\section{Method}

We carried out the three dimensional, radiative hydrodynamic simulations with the JUPITER code \citep{Borro06,Szulagyi15,Szulagyi16} developed by F. Masset \& J. Szul\'agyi. This code is grid-based, solves the continuity and Navier-Stokes equations, the total energy equation and the radiative transfer with flux limited diffusion approximation according to the two-temperature approach \citep[e.g.][]{Kley89,Commercon11}. With the nested meshing technique it is possible to place high resolution meshes around the planet to zoom into its vicinity (similarly to adaptive mesh refinement, but around a single location in the grid), with having the entire circumstellar disk still simulated on the base mesh with lower resolution. This way our maximal resolution in the planet vicinity was $\sim 80\%$ of Jupiter-diameter, roughly 112000 km for a cell-diagonal. This sub-planet resolution allows to examine the circumplanetary gas in unprecedented detail.

Each simulation was made alike, according to the following procedure (similarly to \citealt{Szulagyi16}). First, a Minimum Mass Solar Nebula (i.e. $\sim$11 $\mathrm{M_{Jup}}$) circumstellar disk is simulated between 2.08 AU till 12.40 AU (sampled in 215 cells), with an initial 7.4 degree opening angle (from the midplane to the disk surface, using 20 cells). We applied a constant kinematic viscosity, which equals to 0.004 $\alpha$-viscosity at the planet location. The initial surface density is a power-law function of radius with exponent -0.5 and equals $2222 \rm{kg m^{-2}}$ at the planet's location. The initial disk's aspect ratio is uniform and equal to 0.05. We ran this initial setup for 150 orbits with only two cells in the azimuthal direction ($\pi/2$ each), in order to reach thermal equilibrium in the azimuthally symmetric circumstellar disk. As the heating and cooling effects balance each other, the circumstellar disk evolves and finds a new equilibrium. In the next step, after the thermal equilibrium of the protoplanetary disk has been reached, we re-binned the cells azimuthally for the final resolution (680 cells azimuthally over 2$\pi$) and a point-mass planet was placed at 5.2 AU. This planet was built up to the final mass gradually during 30 orbits. This way the simulation was running with only the base mesh for 150 orbits until we found that the planetary gap profile does not evolve anymore before we began placing the nested meshes around the gas-giant to enhance the resolution. The nested meshes were placed gradually after each other when steady state has been reached on a given level. This usually took a few orbits on each mesh from level 1 till 6. The results were obtained between the 240th and the 250th orbits usually. To avoid singularity at the planet point-mass location, the gravitational potential of the planet was smoothed with usual epsilon-smoothing technique:
\begin{equation}
 U_p=-\frac{G M_p}{\sqrt{x^2+y^2+z^2+{\epsilon}^2}}
\end{equation}
This means that the planet potential ($U_p$) is shallower within a distance of $\epsilon$ in the intermediate vicinity of the planet point mass, i.e. within this short distance the gravity of the Jupiter-mass planet is underestimated. Therefore, it is critical to set the smoothing length ($\epsilon$) as small as possible, but not so small that leads to numerical problems and inaccuracies. Because our resolution doubled on each nested mesh, we set different smoothing lengths on the various levels, few times of cell diagonals on a given grid level, as described in Table \ref{tab:simulations} for the finest level smoothing lengths. At each level of refinement (except the last, 6th one) the smoothing was reduced through a tapering function, described in \citep{Szulagyi14}. For the boundaries and resolution of each refined level, we used the same as Table 1 in \citet{Szulagyi16}. As it was mentioned above, the resolution on the finest level (\#6) was $\sim 80\%$ of Jupiter-diameter ($\sim112000$ km) for a cell-diagonal for all simulations.

The coordinate system was spherical, centered on the star, co-rotating with the planet. The equation of state was ideal gas -- $P=(\gamma-1)\mathrm{E_{int}}$ --  which connects the internal energy ($\mathrm{E_{int}}$) with the pressure (P) through the adiabatic index: $\gamma=1.43$. Thanks to the radiative module and the energy equation, the gas can heat up through viscous heating, adiabatic compression and cool through radiation and adiabatic expansion. For opacities we used the \citet{BL94} opacity table where both the gas and dust opacities are included. This way even though there is no dust explicitly simulated, the dust contribution to the temperature is taken into account through the dust-to-gas ratio. We chose a ratio of 0.01, i.e. equal to the interstellar medium value. The mean-molecular weight was set to 2.3, which corresponds to solar composition.

We carried out 12 different simulations of 1, 3, 5, 10 $\mathrm{M_{Jup}}$ planets, with and without fixing the planetary temperature (Table \ref{tab:simulations}). As it was shown in \citet{Szulagyi16}, with this sub-planet resolution simulations it is possible, and important to parametrize the planet. For this, three basic parameters needs to be fixed: the planetary mass, the planetary radius (equal to two cell-radius, roughly 3 $\mathrm{R_{Jup}}$ for all planetary masses) and the planet temperature. Given that in our simulations the planet parametrized this way consists only of $4^3=64$ cells, the planet temperature was assumed to be homogeneous within this sphere. Regarding choosing the planet temperature values, it is very difficult to estimate the young Jupiter's temperature during its formation, when it was still embedded in hot gas and heavily accreted. We used estimations from planet interior \& evolution models of Mordasini et al. in prep and 
\citet{Guillot95} where the youngest age calculated was at 1 Myr. \citet{Guillot95} provides an effective temperature of 1000 K if Jupiter formed in vacuum, so we chose this planet temperature value as our lowest. The Mordasini et al. models on the other hand took into account that planets form in a background disk and they also have accretional luminosity, thus the effective temperatures of Jupiters range from 2000 K to 8000 K in these calculations for 1 Myr. Hence, for the 1 Jupiter-mass calculations, we chose the upper limit of 10000 K as our hottest planet simulation, and we carried out additional 8000K, 6000K, 4000K, 2000K $\mathrm{T_{planet}}$ computations. As the planet radiates away its formation heat, it gradually cools during this young age, therefore our temperature continuance from 10000 K till 1000 K can be understood as an evolutionary sequence. Regardless what effective temperature Jupiter had initially, it had to pass through at least some of these temperature values, as Jupiter's effective temperature today is only $\sim$ 124 K \citep{Guillot-Jupiterbook}. For the 3-10 $\mathrm{M_{Jup}}$ simulations it is even more difficult to estimate temperature values, given the fewer planetary interior calculations made in the mass regime. Moreover, the simulations are especially time-consuming for the higher planetary masses (due to the deeper gravitational potential of the planet), therefore we could not run such a fine planet temperature grid like in the 1 Jupiter-mass cases. Hence, we made simulations without temperature cap ($T_p > 12000$ K), and with 4000 K, given that a priori, higher mass planets ($\geq 3\mathrm{M_{Jup}}$) should be hotter at the same age, than the Jupiter-mass planets. 

\begin{table}
\centering
  \caption{Simulations}
 \label{tab:simulations}
  \begin{tabular}{cccc}
 \hline
 $M_p$    &   $T_p$    &    \multicolumn{2}{c}{Final smoothing length}  \\
  $[\mathrm{M_{Jup}}]$   &  $[K]$   &  [cell-diagonal] &  [km]\\
 \hline
1 & 10000  & 6 & $6.57\times10^5$ \\
1 & 8000  &  6  & $6.57\times10^5$\\
1 & 6000  & 6  & $6.57\times10^5$ \\
1 & 4000  & 6 & $6.57\times10^5$\\
1 & 2000  &  6 & $6.57\times10^5$\\
1 & 1000  & 6  & $6.57\times10^5$\\
3 & non-fixed  & 12 & $1.31\times10^6$ \\
3 & 4000  & 12 & $1.31\times10^6$  \\
5 & non-fixed      &  12 & $1.31\times10^6$  \\
5 & 4000    & 12 & $1.31\times10^6$  \\
10 & non-fixed    & 24 & $2.63\times10^6$\\
10 & 4000   & 24 &  $2.63\times10^6$\\
 \hline
\end{tabular}
\end{table}

The benefits and disadvantages of using temperature-cap for the planets in the simulations are the followings. First, the surface temperature of forming planets probably should not be higher than 8000 K, because none of the current interior models predict that (e.g. Mordasini et al. submitted). Indeed, 8000 K is much higher than that of the Sun's effective temperature. If the planets were that hot, observations could probably detect them even if these forming planets are embedded. However, there is no observational evidence of that hot-spots of forming planets within the first million year of disk evolution. Secondly, letting the radiative module compute the temperature at the plant location is unlikely correct without a proper planet interior model. The first step toward setting up a planet in the disk simulations is exactly by fixing the temperature,  mass, and the radius of the planet (that is 3 $\mathrm{R_{Jup}}$). Due to the low resolution inside the planet (4 cells radially), a more complex interior model could not be implement, due to the insufficient resolution of planet phase transitions, convective and radiative zones, etc. On the other hand, the temperature-cap is an artificial energy sink, that does not correspond to the local density value. However, along the similar idea, past hydrodynamic simulations often used mass sink cells to enforce/mimic planetary accretion. Considering all the advantages and disadvantages of using a temperature-cap, here we discuss both type of models to show the similarities and differences with and without fixing the planetary temperature.


\section{Results}

\subsection{The Density of the Circumplanetary Gas and the Subdisk Mass}

In this Section we focus on the density distribution of the circumplanetary gas as well as on the mass of the Hill-sphere and the circumplanetary disk. First, the Jupiter-mass simulations with various planetary temperatures are shown, then, the 3-10 Jupiter-mass simulations with and without the planetary temperature cap.

\subsubsection{Subdisk mass and density of the Jupiter-mass planets with various planetary temperature}

Our previous study \citep{Szulagyi16} have showed that the gas temperature can be so high around a Jupiter-mass planet, that the envelope around the gas-giant cannot collapse into a circumplanetary disk. Thus, depending on the temperature, we might get a circumplanetary disk or a circumplanetary envelope, even around giant planets (see also in \citealt{Ormel15a}, \citealt{Ormel15b} and \citealt{DB13}). However, at what exact temperature this transition happens between 2000K and 13000K and how quick is this change, required further simulations with various planet temperature values in between, these are shown here.

\begin{figure*}
\centering
\includegraphics[width=12cm]{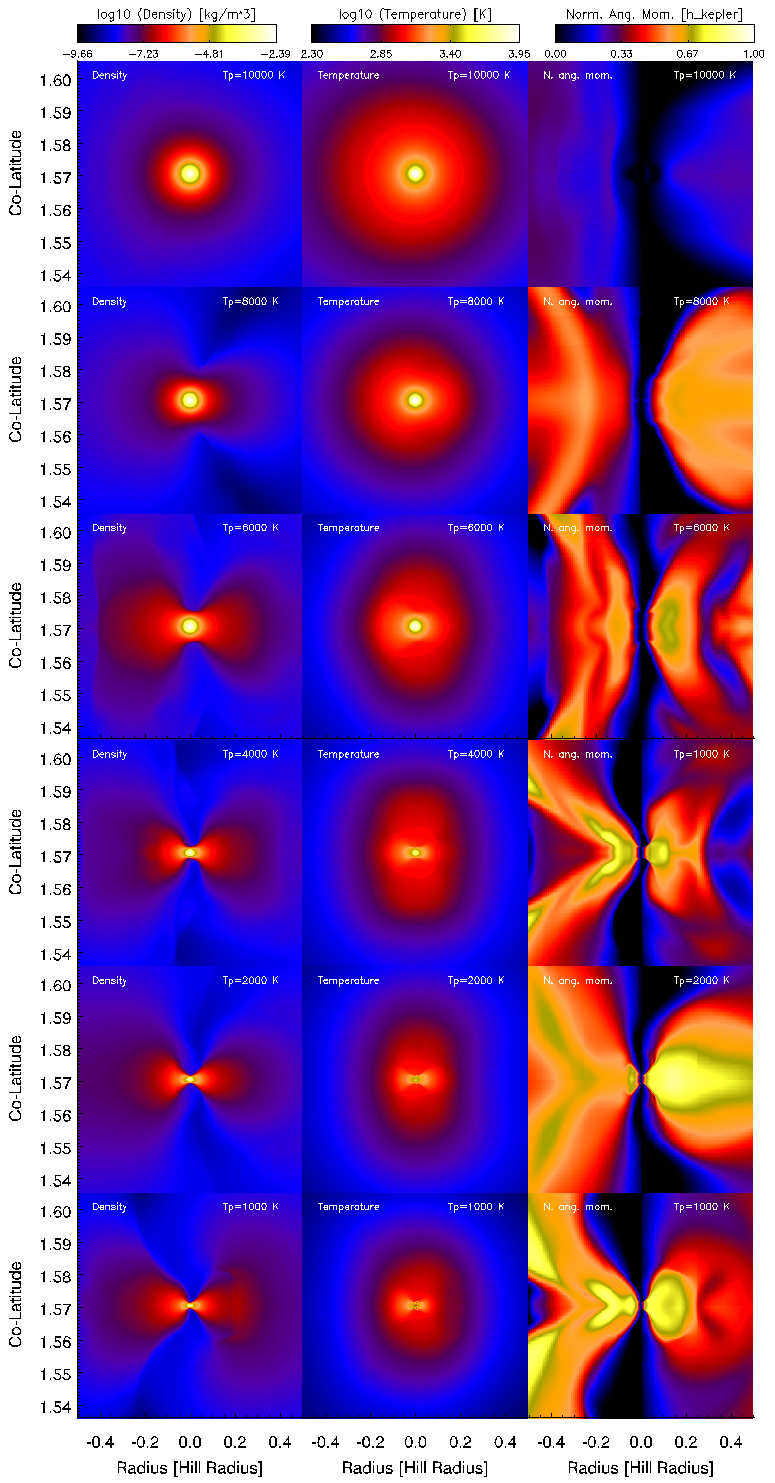} 
\caption{Volume density (first column), temperature (second column), and normalized angular-momentum (third column) color-maps of the planet vicinity.}
\label{fig:pasted}
\end{figure*}

Figure \ref{fig:pasted} shows the density, temperature (see in Sect. \ref{sec:temperature}) and normalized angular momentum (see in Sect. \ref{sec:kinematics}) vertical slices in the planet vicinity for the 6 simulations with various planetary temperatures in increasing order. All color-scales are fixed to a minimum and a maximum value, therefore the color differences and the contrasts for the various simulations can be compared. The figures purposely show only one snapshot in the end of the simulations, in order to see the details of the gas dynamics  due to the high resolution, which would be wiped away by averaging the snapshots in time.

\begin{figure}
\centering
\includegraphics[width=0.8\columnwidth]{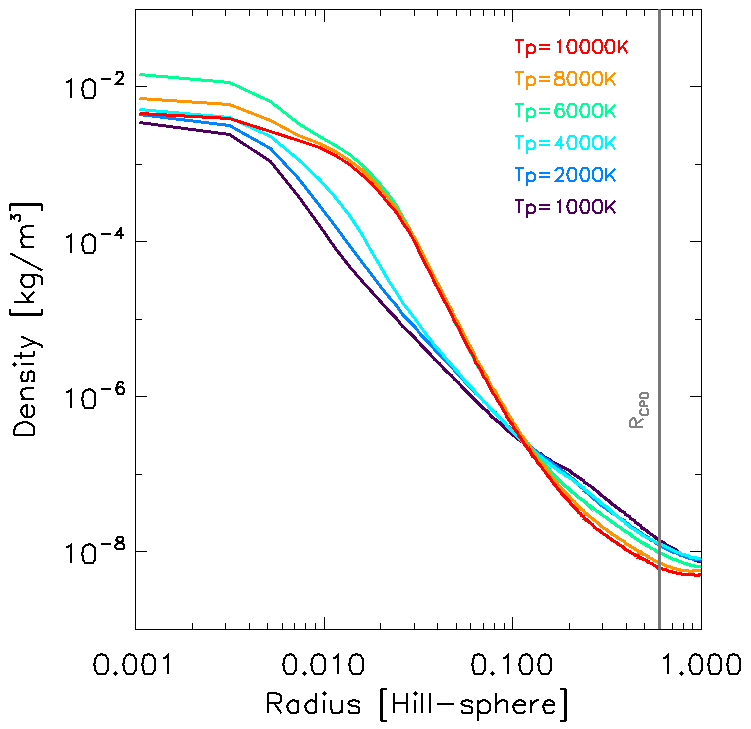} 
\caption{Midplane density-profile in the Hill-sphere of the Jupiter-mass planet with various planetary temperatures from 10000\,K till 1000\,K. The profile is averaged over 21 output of the simulation during one orbit of the planet. There is a clear difference between the simulations of $T_p=$10000-6000 K and the 4000-1000 K cases, hence one can say that the transition from envelope to disk happens between 6000 and 4000 K planetary temperature.}
\label{fig:densprof}
\end{figure}

The first column of Fig. \ref{fig:pasted} represents the density color-maps in decreasing $T_p$ order. We can follow the transition from a circumplanetary envelope to a circumplanetary disk as the temperature of the gas giant decreases. It is a quite continuous transition, therefore, it is difficult to set one point where the disk to envelope transition happens, there are clearly ``disky-envelope" cases based on the density maps. However, with the help of the one dimensional midplane density profile (see Fig. \ref{fig:densprof}) we can set a limit between $T_p$=6000 and $T_p$=4000 K, because the midplane density distribution of the circumplanetary gas is significantly different for the 10000-6000 K $T_p$ models and the 4000-1000 K simulations. These density-profiles are averaged in time over one orbit of the planet (sampled in 21 epochs) and they show the entire Hill-sphere on the midplane, with the planet being on the left-hand side and the edge of the Hill-sphere is on the the right-hand side. From Fig. \ref{fig:densprof} it is clear that the inner parts of the subdisks or envelopes contain the most of the mass of the Hill-sphere. In other words the profile is peaky as we approach the planet.

To compute the mass of the circumplanetary disk, a definition has to be made about what are the limits of the subdisk. In \citet{Szulagyi14} three different methods are described. The first is using the normalized angular momentum of the subdisk (see its detailed definition in Sect. \ref{sec:kinematics}), i.e. setting a minimum threshold for the rotation. In the locally isothermal simulations of \citet{Szulagyi14} the limit was 65\% sub-Keplerian rotation or more, however, as Fig. \ref{fig:pasted} third column shows, as the planet temperature decrease, the rotation of the subdisk enhances. Therefore, there would be difficult to set a minimal normalized angular momentum value which would fit all our simulations. The second method to define the borders of the subdisk could be to compute the eccentricity of a fluid element in respect to the distance to the planet. \citet{Szulagyi14} chose to define the radial extent of the circumplanetary disk (CPD) based on where the orbits of the fluid elements are circular. However, the subdisk can be eccentric \citep{D15,KD06}, therefore this definition can be too restrictive. The third possibility is  drawing the streamlines of the flow around the planet and define the circumplanetary disk where the streamlines bound to the planet. However in this way defining a time-constant 3D surface is a difficult task. A more comprehensive way to define the subdisk borders, therefore, is based on a minimal density, i.e. an isodensity surface. In this work we decided to use this technique with a density threshold of $>1.27\times10^{-8} \mathrm{kg/m^3}$, also taking account where the gas significantly rotating around the planet (see later in Sect. \ref{sec:kinematics}, the first column of Fig. \ref{fig:velopasted}). This value means a $\sim$0.6 $\rm{R_{Hill}}$ radial extent on the midplane, which is consistent with the place where the radial velocity changes sign (see later in Sect. \ref{sec:kinematics} with Fig. \ref{fig:radvelprof}). This value is in agreement with 2D simulations, e.g. \citet{Crida09}. The inner radius of the circumplanetary disk in our calculation was defined at the smoothing length, because the mass within this area would have already been accreted by the planet.

The subdisk masses are shown in Table \ref{tab::Hill-mass}, where the values corresponds to an average value over one orbit of the planet sampled in 21 times, and the standard deviation of these are the errorbars on the average value. Given that the density profiles are so steep (see Fig. \ref{fig:densprof}), it has not a significant importance where (at what minimal density) the outer limit of disk is defined, since the most mass is in the inner subdisk. Due to the fact that the circumplanetary disk boundaries are either way  arbitrary, we also computed the Hill-sphere masses in all simulations (see in Table \ref{tab::Hill-mass}). The Hill-sphere can be defined quite precisely with $\rm{R_{Hill}}=a_p(M_p/3M_*)^{1/3}$ as long as the planet mass is significantly lower than the stellar mass, like in our case (0.001 planet-to-star mass ratio).
In the case of the Hill-sphere mass computation, we integrated the mass of all (entire) cells inside the Hill-sphere, including the mass within the smoothing length, what was not considered for the subdisk mass. From Table \ref{tab::Hill-mass} one can see that there is significant difference between the Hill-masses of the disky cases (1000K $\leq T_p \leq$ 4000K) and the envelope cases (6000K $\leq T_p \leq$ 10000K). The disky cases have a low mass Hill-sphere, roughly $3\times10^{-3} \mathrm{M_{Jup}}$ while the envelope cases almost three times more massive with $\sim8\times10^{-3} \mathrm{M_{Jup}}$. This is most likely a geometric effect: the envelope cases fill up the Hill-sphere spherically, while the disk cases have a low density cone above the circumplanetary disk, where the vertical influx penetrates. 

The subdisk masses for the $T_p=$4000, 2000 and 1000 K simulations are $\sim1.3 \times 10^{-3} \mathrm{M_{Jup}}$, with a slight decline with decreasing planetary temperature, however this cannot be concluded because of the large error-bars. Although, the Hill-sphere mass is also decreasing the same way, supporting this trend. If this is a true relationship, it can be understood again with the geometry: the disks around cooler planets are thinner, the vertical influx's low-density cone is wider, filling up the Hill-sphere less than in the hotter planet simulations. The radial extent of the circumplanetary disks are roughly $0.6 \, \mathrm{R_{Hill}}$, similarly to the findings of \citet{Szulagyi14}, but a bit larger than 0.1-0.3 $\,\mathrm{R_{Hill}}$ of \citet{Tanigawa12} and \citet{AB09b}.

\begin{table}
\centering
  \caption{CPD and Hill-sphere masses for the 1 Jupiter-mass planets}
 \label{tab::Hill-mass}
  \begin{tabular}{ccc}
 \hline
  $T_p$    &   CPD-mass  &  Hill-sphere mass\\
  $[K]$   &   [$10^{-3}\mathrm{M_{Jup}}$]  & [$10^{-3}\mathrm{M_{Jup}}$]\\
 \hline
10000  &  - &  7.11  $\pm$ 0.132   \\
8000  &  - &     8.09 $\pm$  0.199  \\
6000  & -  &    9.11  $\pm$ 0.302  \\
4000  & 1.342  $\pm$  0.324   &   3.49  $\pm$  0.383 \\
2000  &  1.254 $\pm$   0.314    &     3.00   $\pm$ 0.455 \\
1000  &  1.279 $\pm$   0.066    &   2.79  $\pm$  0.192 \\
 \hline
\\
\end{tabular}
\end{table}

\subsubsection{Subdisk mass and density of the 3-10 Jupiter-mass planets}
\label{sec:mass-high}

\begin{figure*}
\centering
\includegraphics[width=13cm]{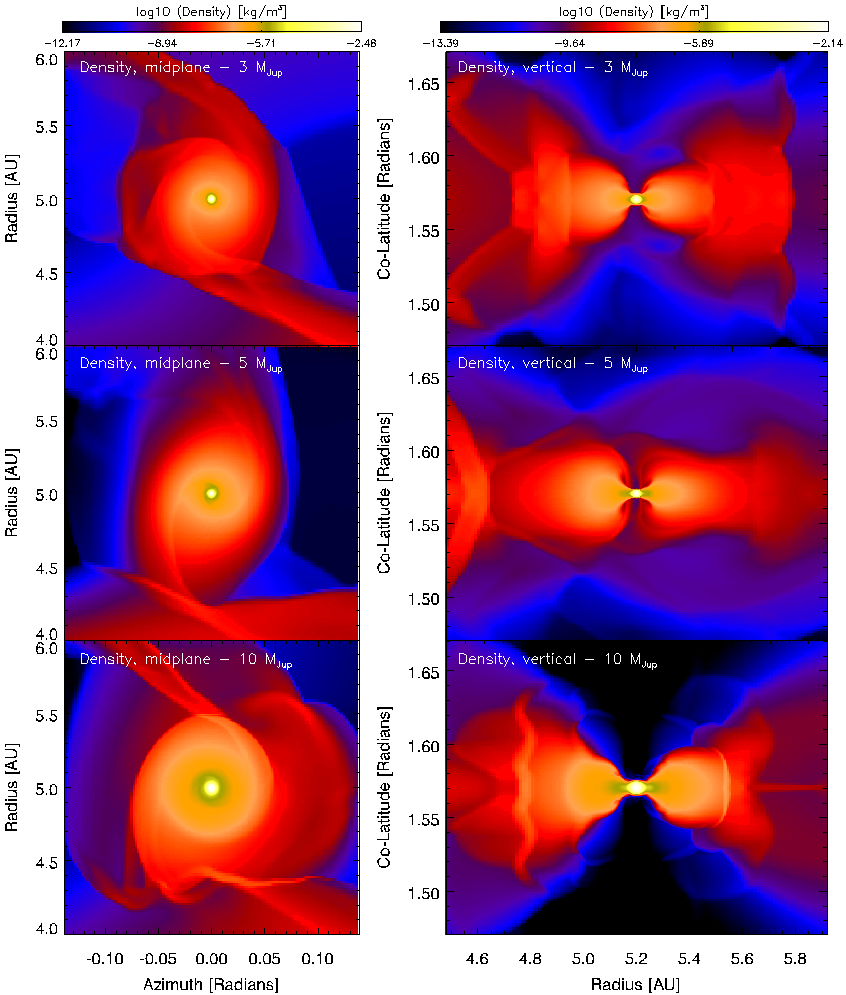} 
\caption{Zoom to the circumplanetary disk face-on (midplane cut, on the left-hand side) and through a vertical slice (right-hand side) for the 3-10 $\mathrm{M_{Jup}}$ planets. In all cases a circumplanetary disk formed around these high-mass gas-giants. The flaring of these disks are larger with higher planetary mass.}
\label{fig:cpd_high}
\end{figure*}

\begin{figure}
\centering
\includegraphics[width=0.8\columnwidth]{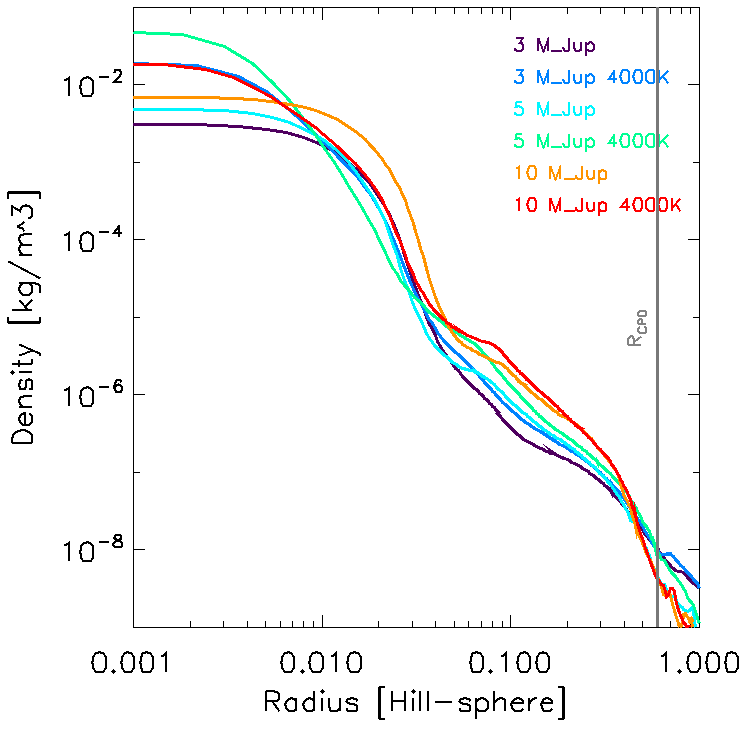}
\caption{Density profile of the midplane for the high-mass planets (3-10 $\mathrm{M_{Jup}}$)}
\label{fig:dens_high}
\end{figure}

In all the high-mass planet cases a circumplanetary disk formed around the gas-giant (see Fig. \ref{fig:cpd_high}). As Figure \ref{fig:dens_high} represents, there is not significant difference between the high-mass planet cases on the mid-plane density of the circumplanetary disks. The mass of the planet seems to be less important to set the density structure than the planetary temperature. The density threshold to define the CPD borders is $>4.225\times 10^{-9}$ kg/m$^3$ in these cases, the radial extent of the disk is again $\sim 0.6 \rm{R_{Hill}}$. The physical size of the CPDs (in AU) are of course significantly different for the different planetary masses, as $\rm{R_{Hill}}$ is diverse too. Regarding the overall mass of the circumplanetary disk, Table \ref{tab::Hill-mass-high} summarizes and Hill-sphere and CPD-masses and their errors. As it can be assumed, the with increasing planetary mass, both the Hill-sphere mass and the CPD-mass is higher. Furthermore, the subdisk mass is always smaller than the Hill-sphere mass, but this is easily understandable, since the CPD is a subset of the Hill-sphere.

\begin{figure*}
\centering
\includegraphics[width=13cm]{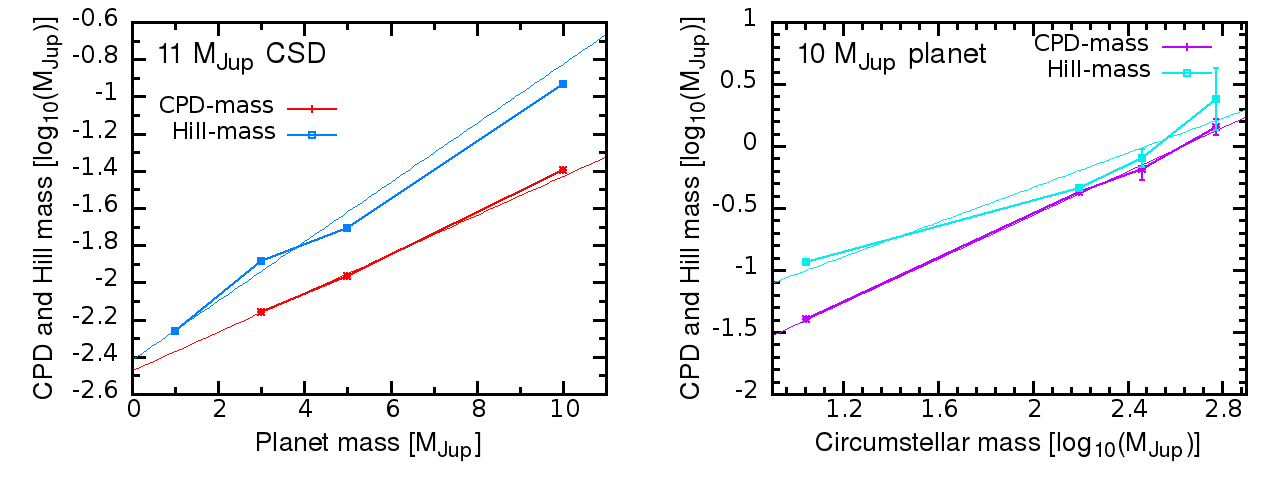}
\caption{Left: Logarithm of circumplanetary disk mass and Hill-sphere mass versus planetary mass for the 11 $\mathrm{M_{Jup}}$ circumstellar disk, and the fitted scaling law (solid thin lines, Eqs. \ref{eq:scaling_mp1}, \ref{eq:scaling_mp2}). Right: Logarithm of CPD and Hill-sphere masses vs. logarithm of circumstellar disk mass and their scaling laws (solid thin lines, Eqs. \ref{eq:scaling_csd1}, \ref{eq:scaling_csd2})}
\label{fig:scaling}
\end{figure*}

With the information on the Hill-sphere and CPD-masses, one can determine the relation for the planet-to-CPD-mass ratio and the planet-to-Hill-sphere-mass ratio. For this, it is important to use only those simulations where the planetary temperature was not fixed. This is because the temperature-cap is an artificial, user defined constraint, which affects the density in the inner CPD (see Fig. \ref{fig:dens_high}). Because the inner CPD accounts for the significant fraction of the entire CPD-mass, any artificial change on the true density should be avoided when calculating such a mass function. Eliminating from Table \ref{tab::Hill-mass-high} those simulations where the temperature was fixed we left with only three values. Using the Hill-sphere masses of the 1 Jupiter-mass planet from \citet{Szulagyi16}, which uses the same setup as the simulations presented in this work, there is now four points to fit the planet-mass to Hill-sphere mass relation (left panel of Fig. \ref{fig:scaling}). Fitting a line to 10-base logarithm values, the mass function is: 

\begin{equation}
\log_{10}(\mathrm{M_{Hill}})=0.159 \times \mathrm{M_{p}}-2.414
\label{eq:scaling_mp1}
\end{equation}

Similarly for the CPD-to-planet mass relation:
\begin{equation}
\log_{10}(\mathrm{M_{CPD}})=0.105 \times \mathrm{M_{p}}-2.473
\label{eq:scaling_mp2}
\end{equation}

\begin{table}
\centering
  \caption{CPD and Hill-sphere masses for the 3-10 $\mathrm{M_{Jup}}$ planets}
 \label{tab::Hill-mass-high}
  \begin{tabular}{cccc}
 \hline
 $M_p$    &   $T_p$    &   CPD-mass  &  Hill-sphere mass\\
  $[\mathrm{M_{Jup}}]$   &  $[K]$   &   [$10^{-3}\mathrm{M_{Jup}}$]  & [$10^{-3}\mathrm{M_{Jup}}$]\\
 \hline
3 & non-fixed (12854) &  6.97 $\pm$    0.24   &   13.08 $\pm$   0.36 \\
3 & 4000  &  6.73  $\pm$  0.17    &  14.73 $\pm$   0.39 \\
5 & non-fixed (19885)     &  10.84 $\pm$   0.44    &  19.70 $\pm$   0.71 \\
5 & 4000    &  23.11 $\pm$   0.53  &  34.39  $\pm$  0.76 \\
10 & non-fixed (16842)   &   40.07 $\pm$    1.18    &   115.99  $\pm$   2.81 \\
10 & 4000   &   29.56  $\pm$  0.38   &  63.06 $\pm$   0.95 \\
 \hline
\end{tabular}
\end{table}

In an earlier work  \citep{SzMayer16} we showed that the CPD-mass scales with the circumstellar disk (CSD) mass too. In that paper, we placed 10 $\mathrm{M_{Jup}}$ planets in various massive circumstellar disk, and we found that the CPD-mass is significantly higher in a significantly more massive protoplanetary disk (see Eq. 3 and 4, and Fig. 4 of that study). With the 10 $\mathrm{M_{Jup}}$ planet simulation from this work, we can add another point to constraint the CPD-to-CSD mass relation (right panel of Fig. \ref{fig:scaling}):

\begin{eqnarray}
\label{eq:scaling_csd1}
\log_{10}(\mathrm{M_{Hill}})=0.70 \log_{10}(\mathrm{M_{CSD}})-1.73\\ 
\log_{10}(\mathrm{M_{CPD}})=0.88 \log_{10}(\mathrm{M_{CSD}})-2.32
\label{eq:scaling_csd2}
\end{eqnarray}

Now that we see the CPD depends on the planet mass and also on the protoplanetary disk mass, we can determine a CPD-mass function based on both $M_{p}$ and $M_{CSD}$. We do this analysis in Sect. \ref{sec:cpd-massfunc}.


\subsection{The circumplanetary disk temperature profile}
\label{sec:temperature}
\subsubsection{Subdisk temperature of the Jupiter-mass planets with various planetary temperature}

We found that the circumplanetary gas temperature is heavily affected by the planetary temperature. In the middle column of Figure \ref{fig:pasted}, one can see that the gas temperature within the Hill-sphere is decreasing as the planet temperature drops. Moreover, again the transition from spherical circumplanetary envelope to disk can be followed as $T_p$ decreases.

\begin{figure}
\centering
\includegraphics[width=0.8\columnwidth]{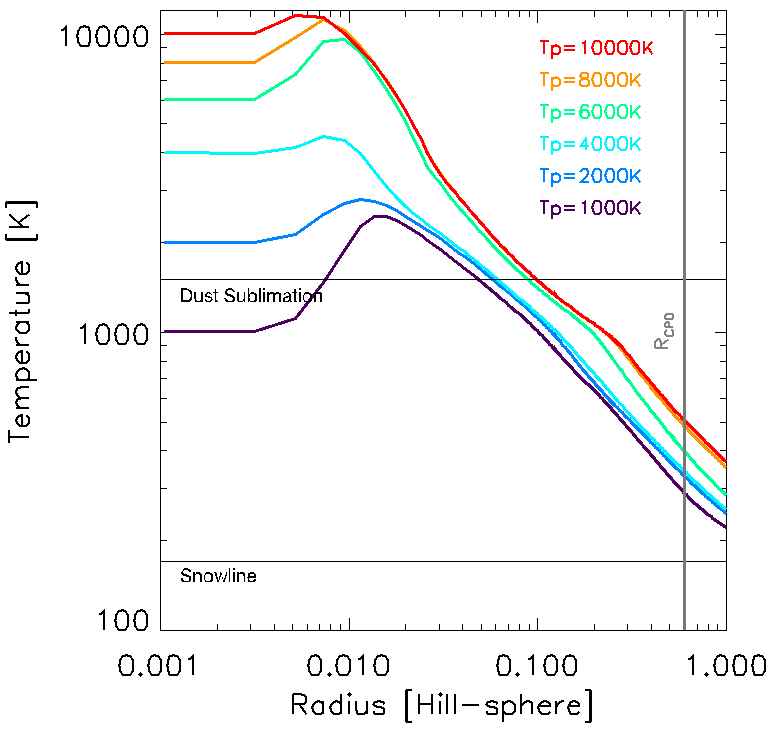} 
\caption{Midplane temperature-profile in the Hill-sphere of the Jupiter-mass planet with various planetary temperatures from 10000\,K till 1000\,K. The profile is averaged over 21 outputs of the simulation during one orbit of the planet.}
\label{fig:tempprof}
\end{figure}

Figure \ref{fig:tempprof} represents the averaged, midplane temperature profiles from the planet (left end of the x-axis) to the Hill-sphere (right end of the x-axis) with logarithmic scaling. The averages are computed again in time, over one orbit of the planet, sampled 21 times. As these temperature profiles show, the gas temperature is heavily affected by $T_{p}$ even far away from the planet, within the {\it{entire}} Hill-sphere. Because the circumplanetary disk is a subset of the Hill-sphere, this means that the gas temperature is lower even beyond the circumplanetary disk borders, as the planet temperature is decreasing. This effect is the largest within $\sim0.02-0.03 \, \mathrm{R_{Hill}}$, but the spread is still within 150 K at 1 $\mathrm{R_{Hill}}$ between the hottest and the coldest simulations. 

In the disky models (simulations of $T_p=$4000K, 2000K, and 1000K), the vertical influx that feeds the CPD, shocks on the surface of the disk, creating a hot ($\sim 4000 K$) and luminous razor-thin shock-front similarly as it was found in e.g. \citet{Szulagyi16} and \citet{SzM16}. The high temperature of this shock can be seen on Fig. \ref{fig:pasted}: the yellow few pixels above the planet on the top layer of the circumplanetary disk. Due to the high-resolution and the shock-capturing method, the shocks are well resolved in the simulations.

In all our simulations the inner circumplanetary disk or -envelope is hotter than the planet itself, similarly to work of \citet{Zhu15}. This is probably an artifact of the temperature cap, as we will see also in Sect. \ref{sec::temp_high}. As photons try to escape from this over-heated region, they could go towards the planet heating that, or they could flee towards the cooler regions of the outer circumplanetary disk. However, due to the temperature cap on the planet region, instead of these photons heating up the planet, we artificially fix the temperature, creating a kind of energy-sink. As we will see in Sect. \ref{sec::temp_high} in the case of 3-10 Jupiter-mass planets, where temperature cap was not set, the planet is always hotter than the circumplanetary disk. 

Given that all in our simulations the circumplanetary disk is hot, it means that it can significantly contribute to the observed luminosity of directly imaged embedded planets. This detection method uses the observed brightness to calculate the luminosity of the planet, which then used to estimate the planetary mass. Often these planets are still surrounded with hot gas \& dust, as observations of thermal emission in the intermediate vicinity of several planets suggest \citep[e.g.][]{Quanz15,Reggiani14,Sallum15}. However, in the luminosity estimation -- therefore in the planetary mass estimation as well -- the contribution of the circumplanetary disk or -envelope is not taken into account (see also in \citealt{Montesinos15}). In this work we showed that this circumplanetary gas is hot at the first 1 Myr of evolution, therefore it is surely contributing to the observed luminosity that might result in overestimation of the mass of directly imaged planets that are still embedded in their protoplanetary disks.

It can also be concluded that the temperature of the circumplanetary gas is heated mostly by the local processes, e.g. adiabatic compression due to the accretion process, irradiation of the forming planet itself, shock-, and viscous heating within the disk. This suggests that the gas in the circumplanetary disk will be hotter and more luminous than the background circumstellar disk in the vicinity, even if the planet is far away from the star. This hot and more extended CPDs at large orbital distances therefore could be easier to detect than the closer in ones. Nevertheless, some heating mechanisms are also decreasing with orbital distance (e.g. the turbulent viscosity), so adequate simulations with large orbital separation planets are needed to study the temperature dependence of CPDs.

The temperature is so high in the circumplanetary gas in all our simulations, that dust condensation (around 1500 K) can only occur beyond 0.04 $\mathrm{R_{Hill}}$ for the $T_p=1000$ K simulation and $>$0.1 $\mathrm{R_{Hill}}$ for the 10000 K planet temperature case. Therefore, as the planet cools off, the inner radius where the temperature is sufficiently low for satellites to form approaches the planet. This suggests that e.g. Jupiter's satellites had to form in the outer circumplanetary disk and migrate inwards. In addition, the $\mathrm{H_2O}$ snow-line is beyond the borders of the circumplanetary disk in all of our models, therefore none of them are sufficient to form the Galilean satellites at this evolutionary stage. The absolute inner radius, where satellites are capable to form is defined by the Roche-limit, within which the satellites would fragment due to tides. Calculating with $R_p=3 R_{Jup}$, a planet-to-satellite density ratio of 0.44, this would give for our Jupiter-mass planet a 0.007 $\mathrm{R_{Hill}}$, which is within the silicate condensation line for all of our models. From all this it can be also assumed that satellite formation happens rather late, when the gas temperature in the subdisk has significantly decreased from these original high values.

Satellites in the gaseous circumplanetary disk can migrate inwards or outwards, depending on the total torque they experience. Depending on whether the proto-satellite is capable to open a gap in the circumplanetary disk, it migrates according to Type II (gap-opening) or Type I (no gap) regime (e.g. \citealt{LP85,Tanaka02,Paardekooper11,BM13}). In our simulations we found the circumplanetary disks to have low gas density and because the turbulent viscosity should be very low \citep{Fujii11,Fujii17} too, the proto-satellites could easily open gaps as long as the disk aspect ratio is not too large. Hence, they most likely would migrate according to the Type II regime \citep{MI16,HI13,Fujii17}. The recent findings of \citet{Fujii17} is that the migration direction of an Io sized body is toward the planet in the bulk of the CPD, but the direction is outward in the outer disk. A self-consistent migration study, that includes satellite growth as well is needed to answer the migration direction and speed in our CPD models, that will be part of a future paper.

Regarding the ionization and viscosity of the subdisk, earlier, locally isothermal work of \citet{Szulagyi14} suggested that the circumplanetary disk might have vanishing viscosity, because planets form in dead zones and the subdisk is shielded from the stellar ionizing photons by the inner circumstellar disk \citep[e.g.][]{Gressel12,Ilgner08,PN10,Baruteau16}. The zero viscosity assumption was likely wrong, as the high temperatures observed in this simulations can ionize the gas at certain places (in the very inner part of the circumplanetary disk, and, in the shock-front on the top layer of the subdisk (see \citealt{SzM16}), which suggests that turbulent viscosity due to the magneto-rotational instability might be at work in such places. However, recent works of \citet{Fujii11} and \citet{Fujii14} argues that it is difficult to imagine that magnetorotational instability takes place in the bulk of the subdisk and drives the accretion as the main angular momentum transport mechanism. Several other instabilities were discovered in circumstellar disks to transport angular momentum \citep{Baruteau14} that can work perhaps similar ways in the circumplanetary disk, such as spiral wave instability \citep{Bae16}, vertical shear instability \citep[e.g.][]{Nelson13,Barker15,Richard16} and disk wind \citep[e.g.][]{Gressel15,Suzuki14,Suzuki10}. For the circumplanetary disk, the spiral wake also thought to promote accretion \citep{Szulagyi14,Zhu16}.

The high temperatures in the planet vicinity can alter the planet migration as well, as numerous works have pointed out \citep[e.g.][]{Benitez15,Benitez16,BM13,Pierens12,Lega14,Lega15,MassetV16}. Likely the planet temperature has even higher effect on satellite migration, as it affects the entire circumplanetary gas.

\subsubsection{Subdisk temperature of the 3-10 Jupiter-mass planets}
\label{sec::temp_high}

\begin{figure}
\centering
\includegraphics[width=0.8\columnwidth]{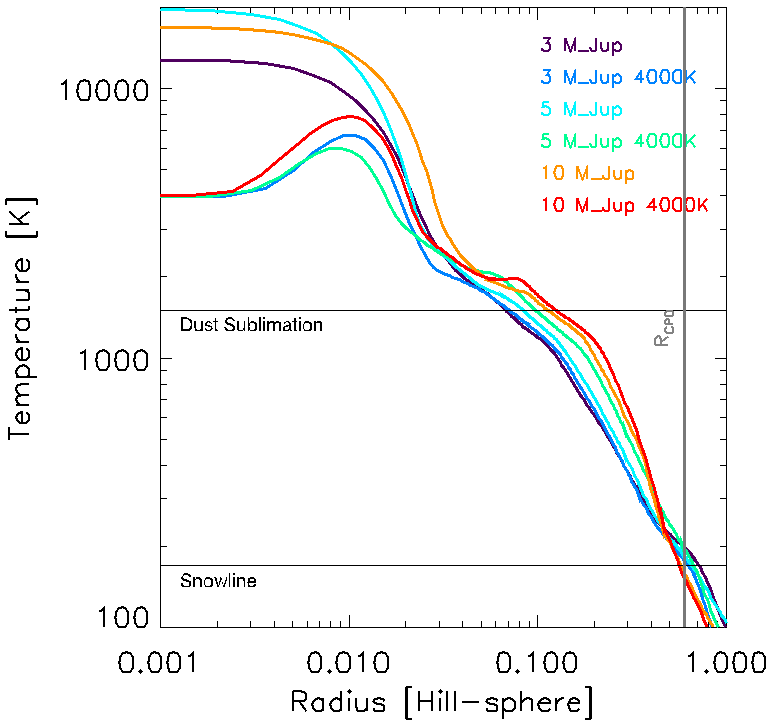} 
\caption{Temperature profiles of the midplane for the high-mass planets (3-10 $\mathrm{M_{Jup}}$). The curves are time-averaged over one orbit of the planet (sampled 21 times).}
\label{fig:temp_high}
\end{figure}

For the 3-10 Jupiter-mass planets, the mid-plane temperature profile (Fig. \ref{fig:temp_high}) the inner 2\% of the Hill-sphere is heavily affected by the planetary temperature, this is less significant in the outer parts of the Hill-sphere. From Fig. \ref{fig:temp_high} it is obvious, that the inner CPD is hotter than the planet in the fixed planetary temperature simulations, but this is not the case when we did not set a temperature cap. This proves that the planet indeed has to be always hotter than its disk, and the inner over-luminous CPD is just an artifact of the temperature fixed simulations. 

Even though these temperature profiles are also averaged over one orbit of the planets sampled in 21 epochs, the curves seem to be much less smooth like in the case of Jupiter-mass planets in the previous sub-section. This is due to the fact that in the case of the high-mass planets the gap, hence the circumplanetary disk also becomes more and more eccentric, and tidally stripped by the protoplanetary disk (see in Sect. \ref{sec::gap_high}). This violent interaction with the circumstellar disk over one orbit of the planet -- while the CPD rotates orders of magnitude times more -- shows on the temperature profiles. 

The location of the dust sublimation line ($\sim$ 1500 K) within the subdisk varies between 0.07-0.11 Hill-radii with increase of planetary mass. This tells us that the satellites can form only in the outer circumplanetary disk, and they have to form even further away as the planetary mass is larger. The water snow-line is just within the CPD borders for the 10 $\mathrm{M_{Jup}}$ planet. It is slightly beyond the subdisk limits for the 3 to 5 $\mathrm{M_{Jup}}$ gas giant. The reason for this that the Hill-sphere of the 10 Jupiter-mass planet is much larger in physical length than the smaller mass planets' Hill-spheres.

The shock-front on the top of the CPDs are much stronger with increasing planetary mass \citep{SzM16}. The ionization of hydrogen can be so high in the shock of the 10 Jupiter-mass case, that $\mathrm{H-\alpha}$ can arise from there. As the gas passes through the shock-front, it loses entropy, so the gas eventually accreted by the planet will have a much lower entropy than without the shock-front on the subdisk \citep{SzM16,OM16}.

\subsection{Kinematic Properties}
\label{sec:kinematics}
\subsubsection{Subdisk velocity and angular momentum of the Jupiter-mass planets with various planetary temperature}

Figure \ref{fig:pasted} third column shows the normalized angular momentum color maps. These are the z-components of the angular momentum relative to the planet and normalized with the local Keplerian rotation. The result is therefore a number between 0.0 and 1.0, where the latter means that the gas is rotating with Keplerian orbital velocity at that location (with $r$ distance from the planet). Usually gas disks are sub-Keplerian, i.e. the gas rotates slower than the local Keplerian rotation at every radius from the planet (or star). On the third column of Fig. \ref{fig:pasted} one can indeed observe that the circumplanetary gas is sub-Keplerian, but rotates faster as the planet temperature drops. This is because it becomes more and more pressure supported as the $T_p$ increases, hence the rotation slows down, and a pressure supported envelope forms. In the $T_p$=10000K case the rotation of the envelope almost stalls. For observational efforts of the circumplanetary disk, all this means that the kinematic fingerprint of the circumplanetary gas (i.e. its rotation) is more obvious if the planet is colder, therefore, it is easier to distinguish the circumplanetary disk kinematically from the surrounding circumstellar disk, similarly as \citet{Perez15} found. In other words, the more evolved (colder) planetary systems are the good targets to detect the subdisk kinematically. 

\begin{figure*}
\centering
\includegraphics[width=12cm]{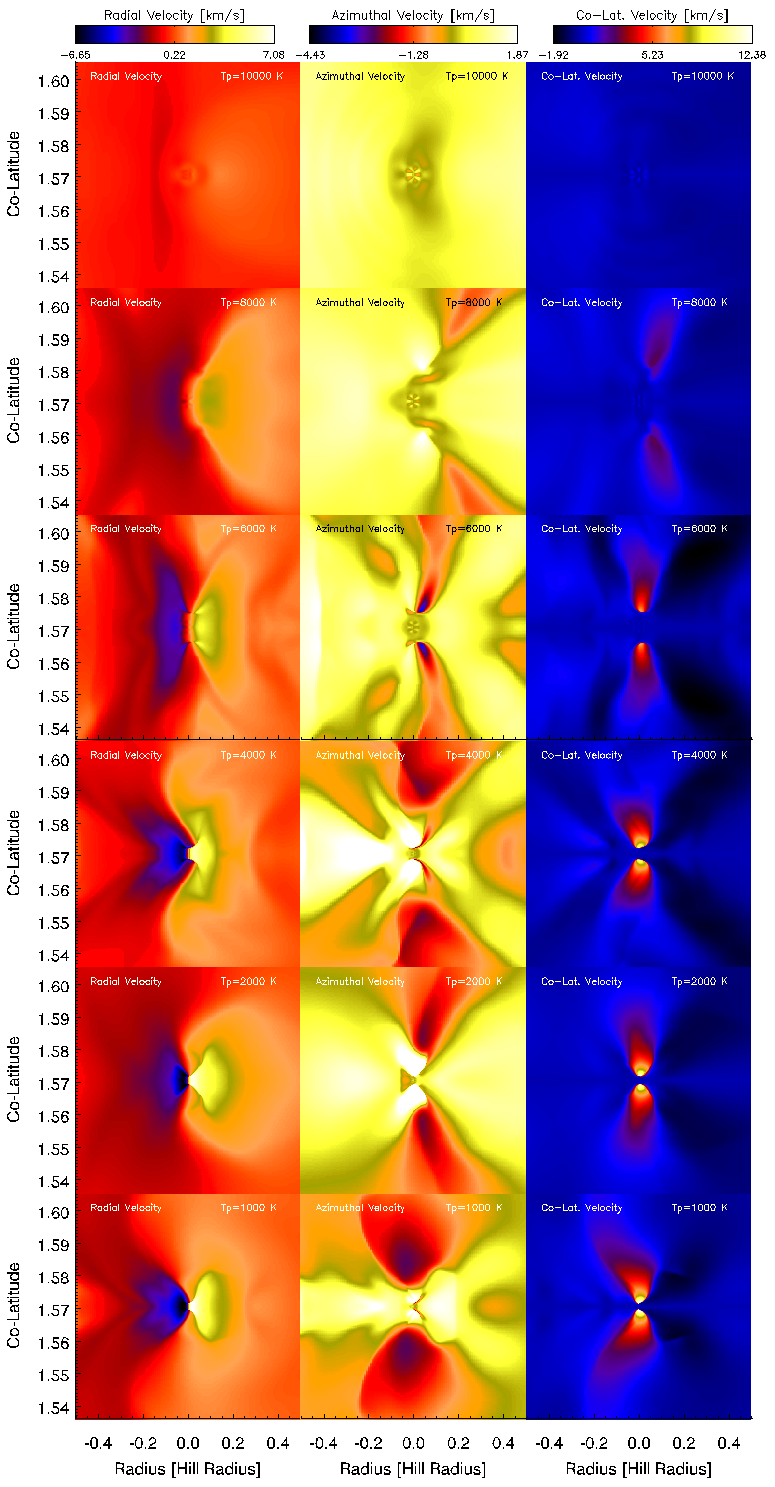} 
\caption{Radial velocity (first column), azimuthal velocity (second column), and co-latitude velocity (third column) color-maps of the planet vicinity.}
\label{fig:velopasted}
\end{figure*}

\begin{figure}
\centering
\includegraphics[width=0.8\columnwidth]{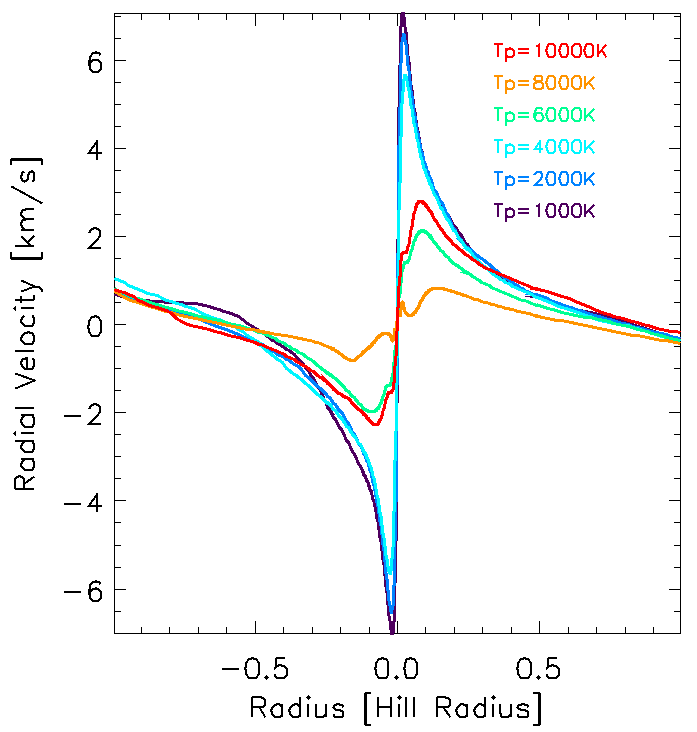} 
\caption{The radial velocity profile in respect to the radius from the star. The radial velocity is defined in respect to the star, not the planet.  }
\label{fig:radvelprof}
\end{figure}

We compare the three different velocity components -- radial, azimuthal and co-latitude velocities -- on Fig. \ref{fig:velopasted} for the different simulations. On the first column one can see the radial velocity defined in respect to the star, therefore the left-side of the planet (that is toward the star) is mainly negative (blue) radial velocity and the right-side of the planet (towards the outer circumstellar disk) is positive (yellow): the CPD rotates into prograde direction. On this figure, with the various planetary temperature it can be already seen that the radial velocity is increasing as the $T_p$ drops. On the 1D averaged radial profiles (see Fig. \ref{fig:radvelprof}) this is even more obvious. The rotation peaks with 7 km/s for the 1000 K planetary temperature model near the gas giant, while the maxima decreases and is located further and further away from the planet for $T_p=2000K$ and $T_p=4000 K$. For the hotter models ($T_p\geq$ 6000K), something different happens: the local maxima of the radial velocity are not in increasing or decreasing order with $T_p$. This is because in these cases we trace convection in the envelope within $\sim 0.2 \mathrm{R_{Hill}}$ as the radial velocity profiles ``wobbles" couple of times (see also in \citealt{Szulagyi16}). The envelopes are too hot, therefore in order to cool, convection starts in the inner parts. Still, the entire envelope is slowly rotating with differential rotation, mostly in prograde direction. The convective motion is beautifully pops out on the second column of Fig. \ref{fig:velopasted}, which represents the azimuthal velocity. Here we can observe a spherical area with checkerboard pattern in the immediate vicinity of the planet within 0.1 $ \mathrm{R_{Hill}}$. This pattern shows the convective cells. As the planetary temperature decreases, the area where convection happens shrinks, however even in the disk cases ($1000K \leq T_p \leq 4000 K$) it can be found. This means that the circumplanetary disk has a substructure; its inner part is always an envelope around the planet with a radius ranging from few Jupiter-radii till few tens of Jupiter-radii depending on the $T_p$, and, within this area, convection could occur. Since between the different simulations only $T_p$ is varied, the size of this envelope is likely set by the pressure due to heating and the gravitational potential of the planet. The balance of these two forces determines where the circumplanetary gas is in the form of an envelope or a disk.

\begin{figure}
\centering
\includegraphics[width=0.8\columnwidth]{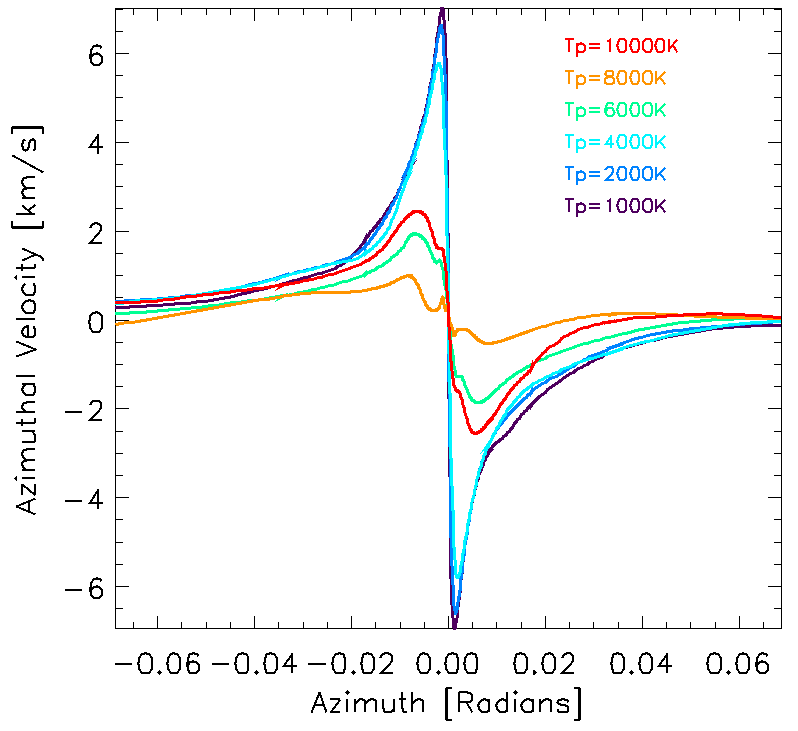} 
\caption{The azimuthal velocity profile versus the azimuth. }
\label{fig:azivelprof}
\end{figure}

The azimuthal velocity profiles (Fig. \ref{fig:azivelprof}) again show the convective wobbles for the 10000K till 6000K planetary temperature simulations and the rotation with 5.5-7 km/s peak velocities for the $T_p=$1000K, 2000K and 4000K models.

\begin{figure}
\centering
\includegraphics[width=0.8\columnwidth]{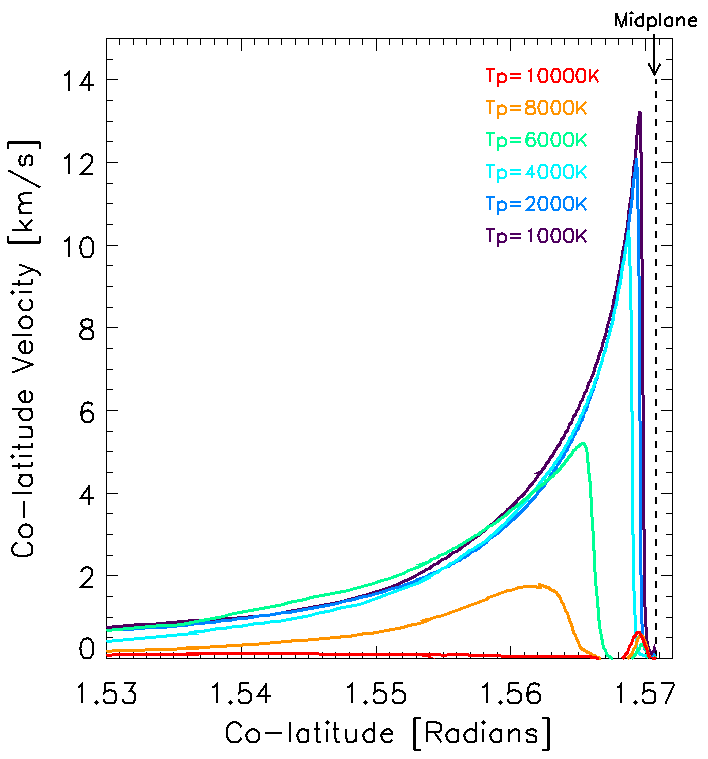} 
\caption{The co-latitude velocity profile in respect to the co-latitude itself. It shows how the vertical influx velocity increases by approaching the planet (which is on right end of the figure on the midplane). In the case of the circumplanetary disks ($1000 K\leq T_p \leq 4000K$), this flow is supersonic, it creates a shock-front on the surface of the circumplanetary disk, here the velocity is the maximal. It can be as high as 10-13 km/s, that might be detectable by the red-shift or broadening of certain spectral lines, which arise from the shock-front. This shock-front gets closer to the planet as the planet temperature decreases and the shock itself is more and more strong, the velocities are higher. In the case of the circumplanetary envelope ($6000 K\leq T_p \leq 10000K$), we do not have shock-front anymore and can only see convection in the envelope as the highest speed motion, therefore the co-latitude velocity is low (0.7 km/s maximum).}
\label{fig:colatvelprof}
\end{figure}

The co-latitude velocity color-maps (third column on Fig. \ref{fig:pasted}) show the velocity of the vertical influx of gas through the planetary gap. Yellow colors show the velocity direction towards the midplane (positive co-latitude velocity). This flow feeds the circumplanetary disk and arising from the top layers of the circumstellar disk (see also in \citep{Szulagyi14}). As it hits the disk, it  creates a shock surface on the top of the subdisk \citep{SzM16}. From this decreasing $T_p$ sequence of Fig. \ref{fig:pasted} it is clear as the vertical influx shocks closer and closer to the planet in the co-latitude direction, while its speed also enhances. In the envelope cases (10000-6000 K planet temperature models) the convection cells again visible in an area within $\sim0.1 \mathrm{R_{Hill}}$. How the shock-front advances toward the planet is also obvious on the co-latitude velocity profiles on Fig. \ref{fig:colatvelprof}. In our hottest model ($T_p=10000$K) the shock-front is not visible anymore at all, here the local maxima on the right-hand side with 0.7 km/s is the convective motion in the envelope. On the other hand, in the coldest case ($T_p=1000$K), the peak velocity of the vertical influx reaches $>$13 km/s, almost the double of the rotational speed. For observational perspective, detecting kinematically the circumplanetary disk (in contrast with the surrounding circumstellar disk's kinematics), the vertical velocities are even better choice than the rotational velocities suggested by \citet{Perez15}. Furthermore, this vertical velocity is so high, that the red-shift or broadening of certain spectral lines might be possible to observe, similarly to photoevaporative flows \citep[e.g.][]{Pascucci11}. 

\subsubsection{Subdisk velocity of the 3-10 Jupiter-mass planets}

\begin{figure}
\centering
\includegraphics[width=0.8\columnwidth]{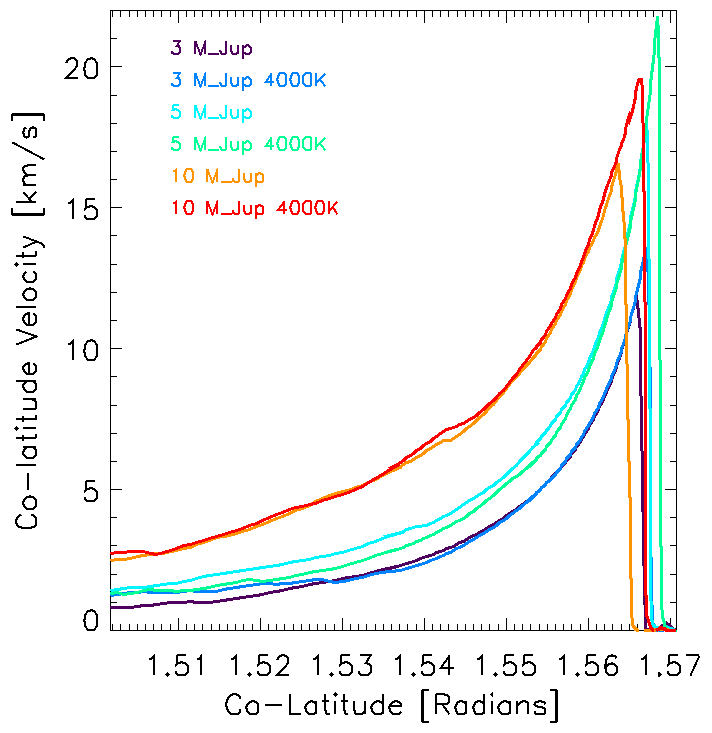} 
\caption{Colatitude velocity profile (versus colatitude) for the high-mass planets (3-10 $\mathrm{M_{Jup}}$)}
\label{fig:colat_high}
\end{figure}

Around this high mass planets, always a circumplanetary disk forms, even if the planetary temperature is over 12000 K. As in the case of the Jupiter-mass planets, the vertical influx shocks closer to the gas giant if the temperature of the planet is lower (Fig. \ref{fig:colat_high}). Apart from the temperature effect, there is of course an effect of the planetary mass -- or rather the choice of the smoothing length of the gravitational potential. This highlights, that for 3D hydrodynamic simulations, the smoothing length can affect the outcome of the simulation, therefore a priori, one should use as small smoothing length as possible. On the other hand, this will slow down enormously the simulations, especially for high planetary masses, like in our case. Better prescription for the softening of the gravitational potential is needed. 

The vertical influx velocity peaks between 12-22 km/s for the 3-10 Jupiter-mass planets. The higher is the velocity as the planetary mass/smoothing length ratio is the largest. The red-shift of certain forbidden lines (such as NeII, NeIII, ArII -- \citealt{Pascucci11}) might be detectable with the current spectrographs, such as VISIR on the VLT. Accretion tracer line emission ($\rm{H\alpha}$, Pa$\beta$, Br$\gamma$) could also be expected, that was already detected in the case of the planetary candidate of LkCa15b by \citet{Sallum15} with Magellan Adaptive Optics System in Simultaneous Differential Imaging mode, but the SPHERE instrument also have the spatial resolution to look for some of these emission lines.

Regarding the rotational velocities, both the radial and azimuthal velocities are higher of course, than in the case of Jupiter-mass planets. The peak values are 12 km/s, 10.5 km/s, 7.6 km/s for the 10, 5, 3 $\mathrm{M_{Jup}}$ planets respectively. 

\subsection{Gap-profile}
\subsubsection{The gaps of the Jupiter-mass planets with various planetary temperature}

\begin{figure}
\centering
\includegraphics[width=0.8\columnwidth]{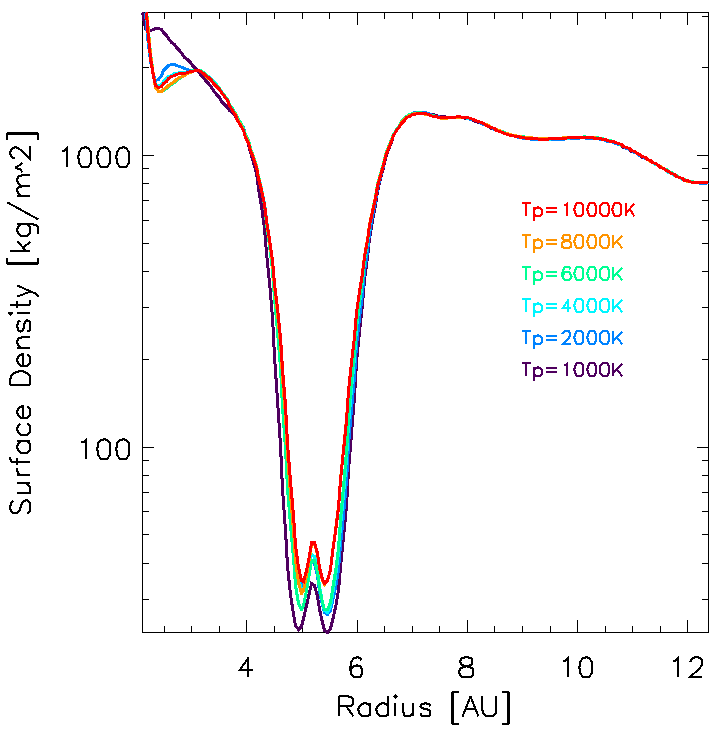} 
\caption{The gap-profile of the different simulations. The temperature of the planet affects the structure of the gap. The colder the planet, the deeper and wider is the gap.}
\label{fig:gap}
\end{figure}

Given that our simulations contain an entire circumstellar disk, we also examined the planetary gap structure for the different simulations. We integrated the volume density in the co-latitude direction on the base mesh, then averaged azimuthally. We averaged the surface density profile in time as well, over one orbit of the planet (sampled 21 times during the orbit). The gap-profiles of the circumstellar disks of the different simulations are shown on Fig. \ref{fig:gap}. We found that the temperature of the planet also affects the gap depth and width. As Fig. \ref{fig:gap} represents, the colder the planet is, the deeper and wider is the gap, although this effect is small. Thus, the planetary temperature is yet again a parameter that somewhat affects the gap properties and was not considered before. This effect is expected, as the hotter the planet, the hotter the gas in its vicinity, therefore the shocks are smoother that results in a shallower, thinner the gap. In Sect. \ref{sec:temperature} we already mentioned that the different planet temperature affects the gas temperature even beyond the the subdisk. In the case of low mass planets, \citet{Benitez15} showed that the planet luminosity alters the torque, which determines the migration of the planet inside the circumstellar disk. In the case of giant planets, the torque also drives the gap-opening, hence the change in planet temperature/luminosity can affect the gap properties.

\begin{figure}
\centering
\includegraphics[width=0.8\columnwidth]{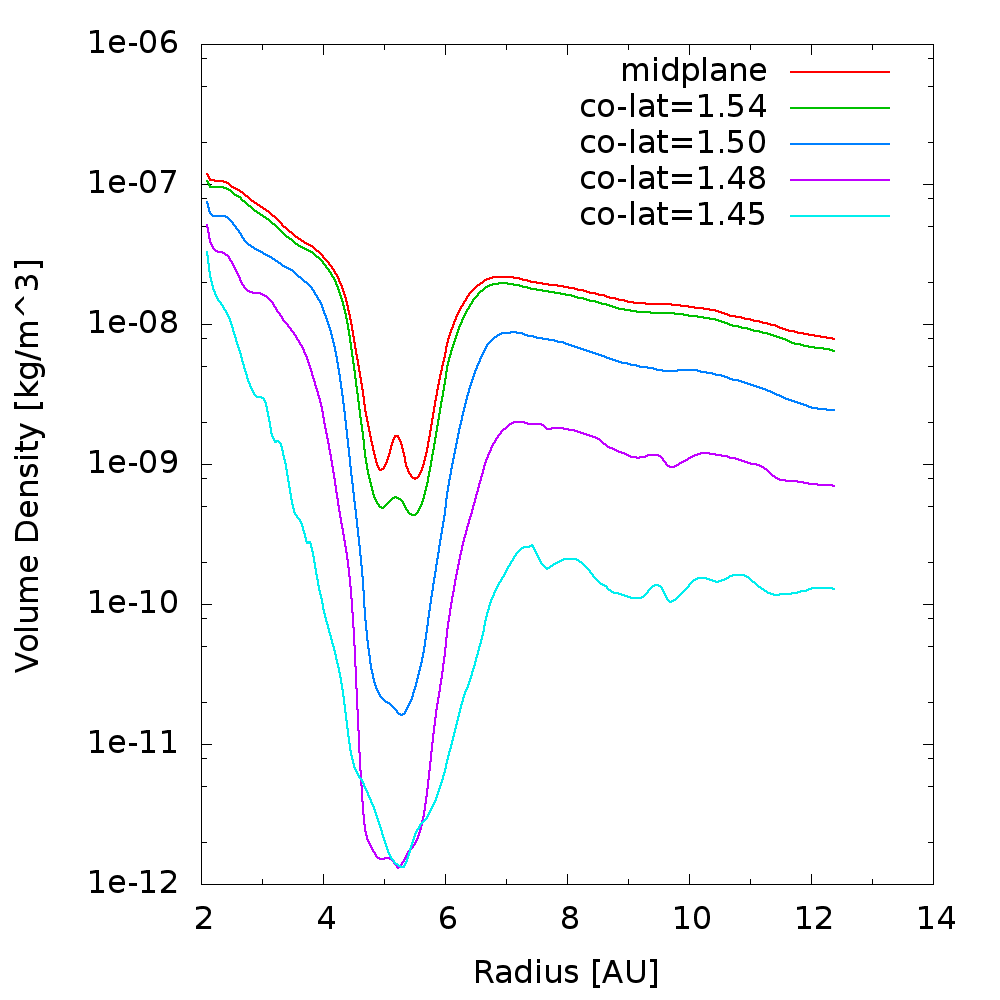} 
\caption{The gap-profile at different co-latitude values (only azimuthally averaged the volume density). The gap depth and width vary while moving away from the midplane; the gap has 3D structure, does not act like 2D gaps.}
\label{fig:gapstru}
\end{figure}

We also examined the structure of the (gas) gap at different co-latitudes. This was motivated by the fact that some works have found that 3D gaps differ from 2D gaps \citep{Kley01,Morbidelli14,MB16}, while others obtained the opposite \citet{Fung16}. Hence, we studied the question by azimuthally averaging the volume density at different co-latitude slices of the circumstellar disk. We show on Fig. \ref{fig:gapstru} how the gap profile changes at different co-latitudes for the $T_p=1000$ K simulation. In agreement with \citet{Morbidelli14}, we also found that the gap is wider while moving away from the midplane in co-latitude direction, hence the gap indeed seems to have a 3D profile and ought to differ from 2D solutions. Moreover, the depth of the gap is also varying with co-latitude distance. On Fig. \ref{fig:gapstru} only one simulation is plotted, but the other simulations qualitatively show the same. The reasons for the difference with some previous works, such as \citet{Fung16} is that we use here radiative simulations, while \citet{Fung16} only locally isothermal equation-of-state. In locally isothermal setup, per definition there is no vertical temperature gradient inside the circumstellar disk, therefore the gap width must be the same at all co-latitudes as it was shown by \citet{MB16} (their Fig. 2 and relevant text). This conservation of the Bernoulli invariant (Eq. 15 in \citealt{MB16}) is the reason why isothermal 2D gap width are the same as isothermal 3D gap widths. However, in radiative simulations -- as could be the case for real circumstellar disk \citep{Woitke09} -- there is a vertical temperature gradient, the Bernoulli invariant is not conserved due to radiative dissipation (cooling through radiation), hence the gap width has to change at different co-latitudes. If one integrate such a 3D disk vertically, the gap will be different depth and width than in a corresponding radiative 2D calculation that traces only the midplane. 

For observational purposes we would like to emphasize that gas gaps differ from dust gaps in depth and width \citep[e.g.][]{Rosotti16,DP15,Tanigawa14}. In this paper we discuss the gas gaps, therefore, only the $\sim$ micron sized dust particles, which are well coupled to the gas, can have the same distribution as the gas gap. Larger particles (mm, cm and beyond) producing usually wider and deeper gaps. In general, even low mass planets can open gaps in the $>$ mm dust \citep[e.g.][]{PM06,Dipierro16}, while only giant planets (beyond $\sim$ Saturn-mass) can open gas gaps.

\subsubsection{Gaps of 3-10 Jupiter-mass planets}
\label{sec::gap_high}

\begin{figure}
\includegraphics[width=\columnwidth]{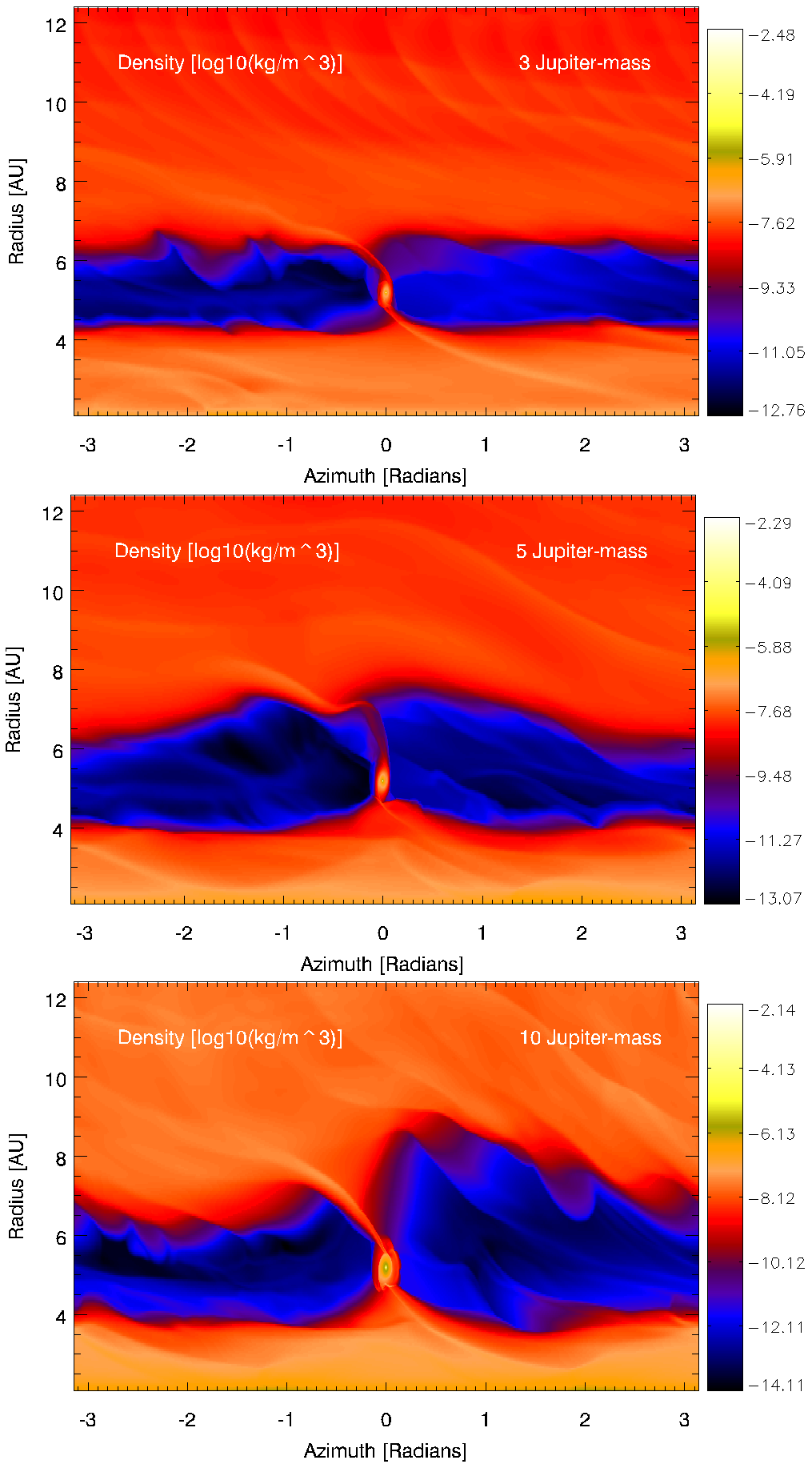} 
\caption{Midplane density maps of the high-mass planets (3-10 $\mathrm{M_{Jup}}$); the gap and the circumplanetary disk is more eccentric with higher planetary mass.}
\label{fig:gap_high}
\end{figure}

\begin{figure*}
\includegraphics[width=17cm]{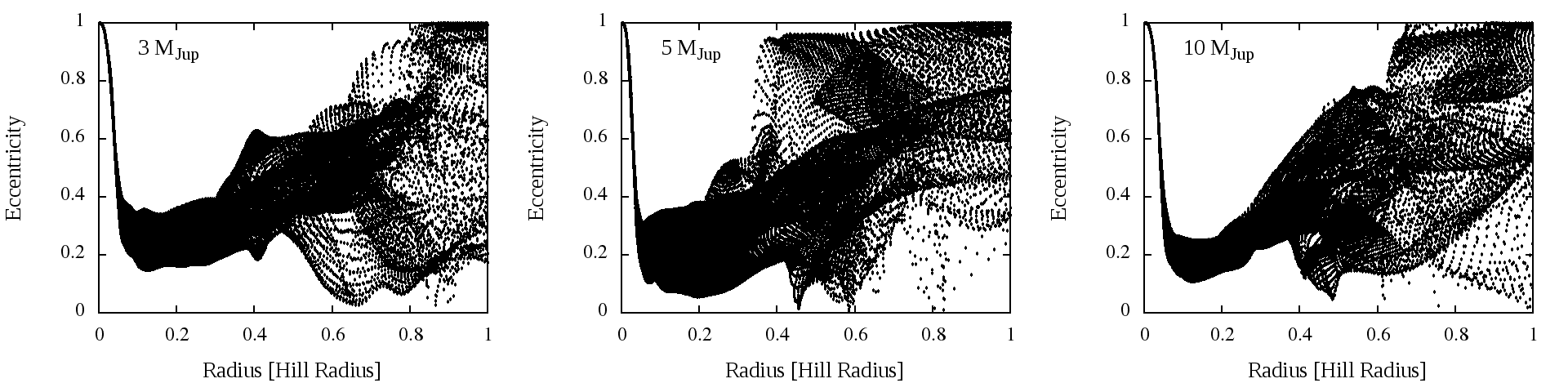} 
\caption{Eccentricity of each cell around the planet on the midplane for the high-mass planets (3-10 $\mathrm{M_{Jup}}$). Clearly, these CPDs around high mass planets are eccentric, and the eccentricity grows with planetary mass.}
\label{fig:ecc}
\end{figure*}

The gap of our 3-10 Jupiter-mass planets are eccentric, as it was shown before by various studies \citep[e.g.][]{Bryden99,Lubow99,Kley99,LDA06,KD06}. It is known, that the eccentricity of the gap is higher with the planetary mass inside the same circumstellar disk, around the same star (see Fig. \ref{fig:gap_high}). Moreover, the circumstellar disk is precessing around the star-planet system and itself becomes eccentric too. This effect is smaller in 3D than in 2D \citep{KD06}, but nevertheless always present. We computed the eccentricity of each cell (fluid element) on the midplane around the planet within the Hill-sphere, these can been seen on Fig. \ref{fig:ecc}. We found that the circumplanetary disk eccentricity is minimal between $\sim$0.1-0.3 Hill-radii, but even in this region the eccentricity varies between 0.1-0.3.

The eccentric CPD means that over one rotation of the circumstellar disk -- which is orders of magnitude more rotation for the circumplanetary disk -- the subdisk changes violently as it is pulled and truncated by the tidal forces. Satellite-formation might be cumbersome in such disks, as long as the circumstellar disk is present. Once the protoplanetary disk dissipates, the perturbation stops, therefore, the satellite-formation is more viable.

\begin{figure}
\includegraphics[width=\columnwidth]{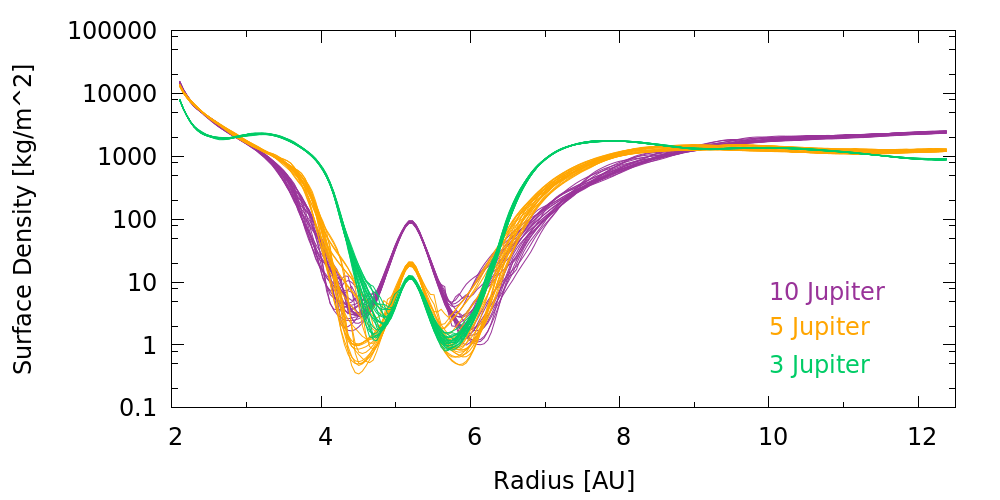} 
\caption{Gap profiles of the high-mass planets (3-10 $\mathrm{M_{Jup}}$) at various times over one orbit of the planet. Due to the eccentric gap, the gap depth and width violently changes at the planet location, but also at different azimuth values.}
\label{fig:gap_high_prof}
\end{figure}

Figure \ref{fig:gap_high_prof} shows the gap-profile of the high-mass gas-giants at various phases during one orbit: clearly the gap width and depth changes rapidly and the more violently with higher planetary mass, similarly to the findings of previous works \citep[e.g.][]{Lubow99,Kley99,LDA06,KD06}. For observations of young planetary systems this means that estimating the planet mass simply from gap depth and width is a very tricky task. Not only that the dust and gas gaps differ significantly for a given mass object, but if the planet has high mass, the gap structure changes rapidly.

\begin{figure}
\centering
\includegraphics[width=0.8\columnwidth]{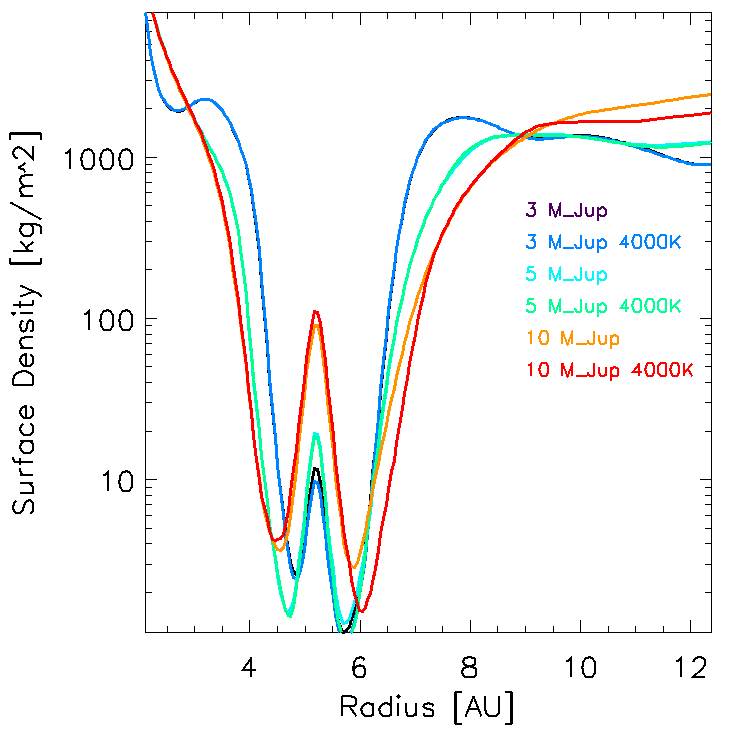}
\caption{Time-averaged gap profiles of the high-mass planets (3-10 $\mathrm{M_{Jup}}$). The gap-profiles are asymmetric and depending on the the temperature of the planet. The lower is $T_p$, the deeper is the gap.}
\label{fig:gapavg_high}
\end{figure}

On Fig. \ref{fig:gapavg_high} we averaged the planetary gap profiles of the 3-10 $\mathrm{M_{Jup}}$ planets over one orbit of the planet. The asymmetry of the left-hand side and right-hand side of the planetary gap is also due to the the eccentricity and precession of the circumstellar disk. The fact that the gap is wider with planetary mass is something which one expects \citep{Duffel15,Fung14,Crida07,Zhu13}. As in the previous section for the Jupiter-mass planet, here again the lower is the temperature of the planet, the deeper is the gap even for the 3-10 $\mathrm{M_{Jup}}$ planet-mass regime. This indicates, that as planets cool off during their evolution, their gap is getting somewhat deeper (not even mentioning that the as the circumstellar disk dissipates the gaps also groove, even more substantially).

\subsection{The Subdisk-mass function}
\label{sec:cpd-massfunc}

We have seen in Sect. \ref{sec:mass-high} that the circumplanetary disk mass scales with the planet mass, i.e. the more massive planets have more massive subdisks in the same protoplanetary disk. Moreover, recently \citet{SzMayer16} showed that the subdisk mass from radiative simulations \footnote{The fact that the CPD surface density scales with the circumstellar disk surface density is trivial in the case of locally isothermal simulations. However, for non-isothermal simulations the form of the scaling law is not anymore that obvious.} also scales with the circumstellar disk mass, because the latter feeds the former. It can also be concluded from Sect. \ref{sec:mass-high} and \citet{SzMayer16} that the circumstellar disk mass is equally important to set the CPD-mass, than the planetary mass. This also means, that it is well possible to find a more massive subdisk around a Jupiter-mass planet in a heavier circumstellar disk, than around a 5 Jupiter-mass planet in a light circumstellar disk. Furthermore, it also indicates that there is a significant mass-evolution of the subdisk as the circumstellar disk vanishes away and therefore the feeding of the CPD is decreasing.

\begin{figure*}
\centering
\includegraphics[width=15cm]{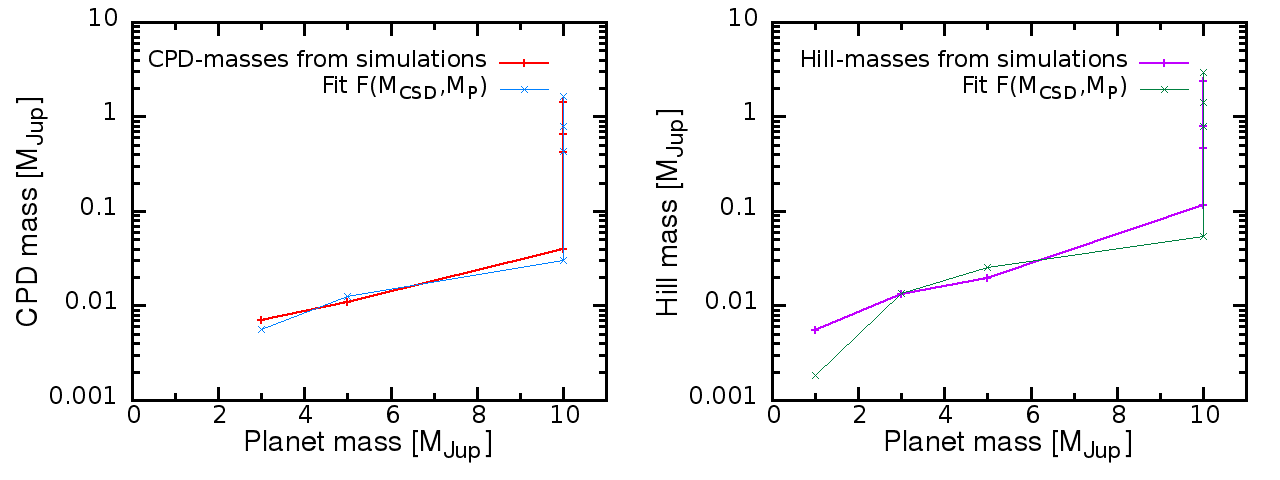}
\caption{The circumplanetary disk (left panel) and Hill-sphere masses (right panel) in a function of planetary mass. The CPD mass depends as strong on the circumstellar disk mass as on the planetary mass. For the fitted functions see Eqs. \ref{eq:mcpd}, \ref{eq:mhill}.}
\label{fig:massfunc}
\end{figure*}

Due to the fact that the CPD mass depends on the planet mass and on the circumstellar disk mass, we used our simulations presented in this paper, in \citet{Szulagyi16}, and in \citet{SzMayer16} to empirically fit a function for $\mathrm{M_{CPD}}= f(\mathrm{M_{CSD}},\mathrm{M_{p}})$. Again, because our definition of circumplanetary disk is arbitrary, this function is also determined for the Hill-sphere mass, which is in contrary a well defined quantity. Because we found that the $\mathrm{M_{CPD}}/\mathrm{M_{CSD}}$ ratio is very small, we estimated the function in the following form: $\mathrm{M_{CPD}}=k \times \mathrm{M_{CSD}} \mathrm{M_{p}}+ l \times \mathrm{M_{CSD}}$ where $k$, $l$ are constants. After fitting to the 7 data-points we have, we obtained a CPD-mass formula:

\begin{eqnarray}
\label{eq:mhill}
\mathrm{M_{Hill}}=5.38\times10^{-4} \mathrm{M_{CSD}} \mathrm{M_{p}}-3.71\times10^{-4}\mathrm{M_{CSD}}\\
\mathrm{M_{CPD}}=3.17\times10^{-4} \mathrm{M_{CSD}} \mathrm{M_{p}}-4.33\times10^{-4}\mathrm{M_{CSD}}
\label{eq:mcpd}
\end{eqnarray}
 
where all masses are in Jupiter-mass unit. We show the data and fitted functions on Fig. \ref{fig:massfunc}.

It is possible to increase the goodness of the fit with higher order polynomials, but given the few data-points and the lack of known physical relation between $M_{CPD}, M_{CSD}, M_{planet}$ we do not feel justified to enter into such a discussion. The above equations can provide useful, though rough estimation of the CPD mass for observation planning.

It is interesting to think about that in case of the Solar System, the integral mass of the satellites of Jupiter and Saturn are constants of $2 \times 10^{-4} \mathrm{M_{p}}$, which implies that the CPD/planet mass ratio could have been constant. However, we see from this work and the results of \citet{SzMayer16}, that for the same massive planet, the CPD mass can be different if the circumstellar disk has a different mass. This means, if such relation exists, it is rather $\mathrm{M_{CPD}}/\mathrm{M_{p}}/\mathrm{M_{CSD}}={\rm constant}$ at the time of the satellite formation. The CPD/planet/CSD mass ratio from all of our simulations gives a small value between 2 and $4\times 10^{-4}$, but of course does not prove a law.

\section{Conclusions \& Discussion}

We performed three dimensional, radiative hydrodynamic simulations of 1, 3, 5, 10 Jupiter-mass planets with sub-planet resolution and varied the temperature of the planet. For the 1 $\mathrm{M_{Jupiter}}$ gas giant we simulated an evolutionary sequence of planet temperatures from 10000 K down to 1000 K representing how the planet radiates away its formation heat. For the higher mass planets (3-10  $\mathrm{M_{Jupiter}}$), we only examined two $T_p$ cases, a non-fixed planetary temperature case, where the temperature peaked above 12000\,K and a fixed 4000 K cap. We examined the circumplanetary disk, or - envelope characteristics based on the planetary temperature. Given that our simulations include an entire circumstellar disk as well, we also studied the planetary gap structure. 

This work is motivated by two reasons. First, we would like to help observational efforts of circumplanetary disks by unveiling their possible mass, density, temperature and kinematic properties. Secondly, the characteristics of the subdisk are also constrain where and when the satellites can form in these disks. 

Whether a planet can form a disk or an envelope depends on the balance between the gravitational forces and the pressure due to heat \citep[e.g.][]{Szulagyi16}. If the latter wins, the outcome is a pressure supported envelope. Our simulations revealed that for 1 Jupiter-mass planet, the transitions from envelope to disk state happens continuously, but mainly between 6000K and 4000 K. For the higher mass planets, we observed circumplanetary disks even if the temperature of the planets were not fixed and rose above 12000 K. As it could be assumed, the average temperature of the subdisk was lower in the simulations when the planet was cooler. Nevertheless, in all cases the entire subdisk temperature was over the water freezing point ($\sim$ 180 K), and the inner 10\% of the Hill-sphere even above the silicate sublimation point. This suggests that satellite formation cannot happen at this early evolutionary stage ($\lesssim$ 1 Myr) of the circumplanetary disk, only much later. The temperature-profiles of the disks were always very steep, suggesting that the conditions for satellite-formation are first adequate in the outer circumplanetary disk; satellites form there and migrate in later on.

The high temperature of the circumplanetary disks suggests high luminosity, similar to how \citet{Zhu15} and \citet{SzM16} found. This is important for the direct detection of young, embedded planets, because their planetary mass estimation is based on the observed luminosity, where the subdisk contribution is not taken into account. This can lead to over-estimation of planetary masses from direct imaging observations of embedded planets.

The circumplanetary disk mass was found to scale with the planet mass, accounting for $\sim1.3,~7,~20,~40 \times10^{-3} \mathrm{M_{Jupiter}}$ for the 1, 3, 5, 10 Jupiter-mass planets respectively. The entire Hill-spheres contained less than three times more mass in all cases. The circumplanetary disk masses are small in respect to the few Jupiter-mass planets, however they are at least an order of magnitude higher than locally isothermal simulations found in the past \citep{DA03,Gressel13,Szulagyi14}. Therefore, the inclusion of thermal processes into the hydrodynamic simulations are important for unveiling the basic characteristics of the subdisk. The disk mass not only scales with the planet, but also with the circumstellar disk in case of radiative simulations too \citep{SzMayer16}. With the circumplanetary- and Hill-sphere masses of this work and \citet{SzMayer16} we constructed an empirical formula to estimate the subdisk mass based on the protoplanetary disk mass and the planet mass (Eqs. \ref{eq:mhill}, \ref{eq:mcpd}), which can help for future observational proposals.

Regarding the kinematic properties of the circumplanetary gas, the envelope cases ($T_p=$10000 K, 8000 K, 6000 K simulations for the 1 $\mathrm{M_{Jupiter}}$) have slow rotation and mostly characterized by convection in the inner parts. As the planetary temperature drops, the rotation enhances, as well as the vertical influx's speed (peaking at 13 km/s for $T_p=$1000 K). For the 3-10  $\mathrm{M_{Jupiter}}$, these velocities are even higher because of the larger planetary mass. For lower planetary temperatures, the vertical influx velocity peaking over 25 km/s for the high-mass planet simulations. This is a very high vertical velocity in respect to the other parts of the circumstellar disk and the gap, which makes it possible to aim for a detection of the subdisk based on its kinematic contrast with the surrounding protoplanetary disk.

Because our simulations contain an entire protoplanetary disk, the planetary gap structure was examined as well. We found that the gap is deeper and wider as the planet radiates away its formation heat, however this effect is relatively small in comparison to other effects (such as change in viscosity). We also showed that the gap has a 3D structure, its profile changes with the vertical distance from the planet, therefore 3D gaps do differ from 2D simulations' gaps. Moreover, we confirmed \citet{KD06} findings that for high-mass planets ($\gtrsim 5 \mathrm{M_{Jupiter}}$) the gap eccentricity increases, hence the circumstellar disk regularly engulfs then truncates the subdisk, creating too violent environment for satellites to form there. With the growing gap eccentricity, the circumplanetary disk becomes more elongated.

To provide more in-depth observational constraints and to create synthetic images for observations from this simulations, a wavelength dependent radiative transfer software is needed to be applied on the hydrodynamic fields. This will be part of a future work.

\begin{acknowledgements}
      
I thank for the anonymous referee the thoughtful review, which helped to improve the paper. Furthermore, I am grateful for the useful discussions with F. Masset, M. R. Meyer and C. Mordasini and for the language editing for S. March. I acknowledge the support from the ETH Post-doctoral Fellowship from the Swiss Federal Institute of Technology (ETH Z\"urich). This work has been in part carried out within the frame of the National Centre for Competence in Research  ``PlanetS"  supported by  the  Swiss  National Science Foundation. Computations have been done on the ``M\"onch" machine hosted at the Swiss National Computational Centre.

\end{acknowledgements}


\begin{thebibliography}{}
\bibitem[Ayliffe \& Bate(2009a)]{AB09a} Ayliffe, B.~A., \& Bate, M.~R.\ 2009, MNRAS, 393, 49 
\bibitem[Ayliffe \& Bate(2009b)]{AB09b} Ayliffe, B.~A., \& Bate, M.~R.\ 2009, MNRAS, 397, 657 
\bibitem[Bae et al.(2016)]{Bae16} Bae, J., Nelson, R.~P., Hartmann, L., \& Richard, S.\ 2016, arXiv:1607.01792 
\bibitem[Barker \& Latter(2015)]{Barker15} Barker, A.~J., \& Latter, H.~N.\ 2015, \mnras, 450, 21 
\bibitem[Baruteau et al.(2014)]{Baruteau14} Baruteau, C., Crida, A., Paardekooper, S.-J., et al.\ 2014, Protostars and Planets VI, 667 
\bibitem[Baruteau \& Masset(2013)]{BM13} Baruteau, C., \& Masset, F.\ 2013, Lecture Notes in Physics, Berlin Springer Verlag, 861, 201  
\bibitem[Baruteau et al.(2016)]{Baruteau16} Baruteau, C., Bai, X., Mordasini, C., \& Molli{\`e}re, P.\ 2016, \ssr,  
\bibitem[Bell \& Lin(1994)]{BL94} Bell, K.~R., \& Lin, D.~N.~C.\ 1994, ApJ, 427, 987 
\bibitem[Ben{\'{\i}}tez-Llambay et al.(2015)]{Benitez15} Ben{\'{\i}}tez-Llambay, P., Masset, F., Koenigsberger, G., \& Szul{\'a}gyi, J.\ 2015, \nat, 520, 63 
\bibitem[Ben{\'{\i}}tez-Llambay et al.(2016)]{Benitez16} Ben{\'{\i}}tez-Llambay, P., Ramos, X.~S., Beaug{\'e}, C., \& Masset, F.~S.\ 2016, \apj, 826, 13 
\bibitem[Brittain et al.(2014)]{Brittain14} Brittain, S.~D., Carr, J.~S., Najita, J.~R., Quanz, S.~P., \& Meyer, M.~R.\ 2014, \apj, 791, 136 
\bibitem[Bryden et al.(1999)]{Bryden99} Bryden, G., Chen, X., Lin, D.~N.~C., Nelson, R.~P., \& Papaloizou, J.~C.~B.\ 1999, \apj, 514, 344 
\bibitem[Canup \& Ward(2002)]{CW02} Canup, R.~M., \& Ward, W.~R.\ 2002, AJ, 124, 3404 
\bibitem[Canup \& Ward(2006)]{CW06} Canup, R.~M., \& Ward, W.~R.\ 2006, Nature, 441, 834 
\bibitem[Commer{\c c}on et al.(2011)]{Commercon11} Commer{\c c}on, B., Teyssier, R., Audit, E., Hennebelle, P., \& Chabrier, G.\ 2011, A\&A, 529, A35
\bibitem[Crida et al.(2009)]{Crida09} Crida, A., Baruteau, C., Kley, W., \& Masset, F.\ 2009, \aap, 502, 679 
\bibitem[Crida \& Morbidelli(2007)]{Crida07} Crida, A., \& Morbidelli, A.\ 2007, \mnras, 377, 1324 
\bibitem[D'Angelo et al.(2003)]{DA03} D'Angelo, G., Henning, T., \& Kley, W.\ 2003, ApJ, 599, 548 
\bibitem[D'Angelo \& Bodenheimer(2013)]{DB13} D'Angelo, G., \& Bodenheimer, P.\ 2013, \apj, 778, 77
\bibitem[D'Angelo et al.(2014)]{DA14} D'Angelo, G., Weidenschilling, S.~J., Lissauer, J.~J., \& Bodenheimer, P.\ 2014, \icarus, 241, 298 
\bibitem[D'Angelo \& Podolak(2015)]{DP15} D'Angelo, G., \& Podolak, M.\ 2015, \apj, 806, 203 
\bibitem[de Val-Borro et al.(2006)]{Borro06} de Val-Borro, M., Edgar, R.~G., Artymowicz, P., et al.\ 2006, MNRAS, 370, 529
\bibitem[Dipierro et al.(2016)]{Dipierro16} Dipierro, G., Laibe, G., Price, D.~J., \& Lodato, G.\ 2016, \mnras, 459, L1
\bibitem[Duffell(2015)]{Duffel15} Duffell, P.~C.\ 2015, \apjl, 807, L11 
\bibitem[Dunhill(2015)]{D15} Dunhill, A.~C.\ 2015, \mnras, 448, L67
\bibitem[Estrada et al.(2009)]{Estrada09} Estrada, P.~R., Mosqueira, I., Lissauer, J.~J., D'Angelo, G., \& Cruikshank, D.~P.\ 2009, Europa, Edited by Robert T.~Pappalardo, William B.~McKinnon, Krishan K.~Khurana ; with the assistance of Ren{\'e} Dotson with 85 collaborating authors.~University of Arizona Press, Tucson, 2009.~The University of Arizona space science series ISBN: 9780816528448, p.27, 27 
\bibitem[Fujii et al.(2011)]{Fujii11} Fujii, Y.~I., Okuzumi, S., \& Inutsuka, S.-i., 2011, ApJ, 743, 53 
\bibitem[Fujii et al.(2014)]{Fujii14} Fujii, Y.~I., Okuzumi, S., Tanigawa, T., \& Inutsuka, S.-i.\ 2014, ApJ, 785, 101
\bibitem[Fujii et al.(2017)]{Fujii17} Fujii, Y.~I., Kobayashi, H., Takahashi, S.~Z., \& Gressel, O.\ 2017, \aj, 153, 194 
\bibitem[Fung et al.(2014)]{Fung14} Fung, J., Shi, J.-M., \& Chiang, E.\ 2014, \apj, 782, 88 
\bibitem[Fung \& Chiang(2016)]{Fung16} Fung, J., \& Chiang, E.\ 2016, arXiv:1606.02299 
\bibitem[Galvagni et al.(2012)]{Galvagni12} Galvagni, M., Hayfield, T., Boley, A., et al.\ 2012, MNRAS, 427, 1725 
\bibitem[Gressel et al.(2012)]{Gressel12} Gressel, O., Nelson, R.~P., \& Turner, N.~J.\ 2012, \mnras, 422, 1140 
\bibitem[Gressel et al.(2013)]{Gressel13} Gressel, O., Nelson, R.~P., Turner, N.~J., \& Ziegler, U.\ 2013, ApJ, 779, 59 
\bibitem[Gressel et al.(2015)]{Gressel15} Gressel, O., Turner, N.~J., Nelson, R.~P., \& McNally, C.~P.\ 2015, \apj, 801, 84 
\bibitem[Guillot et al.(1995)]{Guillot95} Guillot, T., Chabrier, G., Gautier, D., \& Morel, P.\ 1995, ApJ, 450, 463 
\bibitem[Guillot et al.(2004)]{Guillot-Jupiterbook} Guillot, T., Stevenson, D.~J., Hubbard, W.~B., \& Saumon, D., 2004, Jupiter.~The Planet, Satellites and Magnetosphere, 35 
\bibitem[Hasegawa \& Ida(2013)]{HI13} Hasegawa, Y., \& Ida, S.\ 2013, \apj, 774, 146 
\bibitem[Ilgner \& Nelson(2008)]{Ilgner08} Ilgner, M., \& Nelson, R.~P.\ 2008, \aap, 483, 815 
\bibitem[Kley(1989)]{Kley89} Kley, W.\ 1989, A\&A, 208, 98 
\bibitem[Kley(1999)]{Kley99} Kley, W.\ 1999, \mnras, 303, 696 
\bibitem[Kley et al.(2001)]{Kley01} Kley, W., D'Angelo, G., \& Henning, T.\ 2001, \apj, 547, 457 
\bibitem[Kley \& Dirksen(2006)]{KD06} Kley, W., \& Dirksen, G.\ 2006, A\&A, 447, 369 
\bibitem[Kraus \& Ireland(2012)]{KI12} Kraus, A.~L., \& Ireland, M.~J.\ 2012, ApJ, 745, 5 
\bibitem[Lega et al.(2015)]{Lega15} Lega, E., Morbidelli, A., Bitsch, B., Crida, A., \& Szul{\'a}gyi, J.\ 2015, \mnras, 452, 1717 
\bibitem[Lega et al.(2014)]{Lega14} Lega, E., Crida, A., Bitsch, B., \& Morbidelli, A.\ 2014, \mnras, 440, 683
\bibitem[Lin \& Papaloizou(1985)]{LP85} Lin, D.~N.~C., \& Papaloizou, J.\ 1985, Protostars and Planets II, 981 
\bibitem[Lubow et al.(1999)]{Lubow99} Lubow, S.~H., Seibert, M., \& Artymowicz, P.\ 1999, \apj, 526, 1001 
\bibitem[Lubow \& D'Angelo(2006)]{LDA06} Lubow, S.~H., \& D'Angelo, G.\ 2006, \apj, 641, 526 
\bibitem[Masset \& Ben{\'{\i}}tez-Llambay(2016)]{MB16} Masset, F.~S., \& Ben{\'{\i}}tez-Llambay, P.\ 2016, \apj, 817, 19 
\bibitem[Masset \& Velasco(2016)]{MassetV16} Masset, F.~S., \&  Velasco Romero, D.~A. \ 2016, accepted at MNRAS, arXiv:1611.05684 
\bibitem[Miguel \& Ida(2016)]{MI16} Miguel, Y., \& Ida, S.\ 2016, \icarus, 266, 1 
\bibitem[Montesinos et al.(2015)]{Montesinos15} Montesinos, M., Cuadra, J., Perez, S., Baruteau, C., \& Casassus, S.\ 2015, \apj, 806, 253 
\bibitem[Morbidelli et al.(2014)]{Morbidelli14} Morbidelli, A., Szul{\'a}gyi, J., Crida, A., et al.\ 2014, Icarus, 232, 266 
\bibitem[Mosqueira \& Estrada(2003a)]{ME03a} Mosqueira, I., \& Estrada, P.~R.\ 2003, Icarus, 163, 198
\bibitem[Mosqueira \& Estrada(2003b)]{ME03b} Mosqueira, I., \& Estrada, P.~R.\ 2003, Icarus, 163, 232 
\bibitem[Nelson et al.(2013)]{Nelson13} Nelson, R.~P., Gressel, O., \& Umurhan, O.~M.\ 2013, \mnras, 435, 2610 
\bibitem[Ormel et al.(2015a)]{Ormel15a} Ormel, C.~W., Kuiper, R., \& Shi, J.-M.\ 2015, \mnras, 446, 1026 
\bibitem[Ormel et al.(2015b)]{Ormel15b} Ormel, C.~W., Shi, J.-M., \& Kuiper, R.\ 2015, \mnras, 447, 3512 
\bibitem[Owen \& Menou(2016)]{OM16} Owen, J.~E., \& Menou, K.\ 2016, \apjl, 819, L14 
\bibitem[Paardekooper \& Mellema(2006)]{PM06} Paardekooper, S.-J., \& Mellema, G.\ 2006, \aap, 453, 1129 
\bibitem[Paardekooper et al.(2011)]{Paardekooper11} Paardekooper, S.-J., Baruteau, C., \& Kley, W.\ 2011, \mnras, 410, 293 
\bibitem[Papaloizou \& Nelson(2005)]{PN05} Papaloizou, J.~C.~B., \& Nelson, R.~P.\ 2005, \aap, 433, 247 
\bibitem[Pascucci et al.(2011)]{Pascucci11} Pascucci, I., Sterzik, M., Alexander, R.~D., et al.\ 2011, \apj, 736, 13 
\bibitem[Perez et al.(2015)]{Perez15} Perez, S., Dunhill, A., Casassus, S., et al.\ 2015, ApJL, 811, L5 
\bibitem[Pierens \& Nelson(2010)]{PN10} Pierens, A., \& Nelson, R.~P.\ 2010, \aap, 520, A14 
\bibitem[Pierens et al.(2012)]{Pierens12} Pierens, A., Baruteau, C., \& Hersant, F.\ 2012, \mnras, 427, 1562 
\bibitem[Quanz et al.(2015)]{Quanz15}  Quanz, S.~P., Amara, A., Meyer, M.~R., et al.\ 2015, ApJ, 807, 64 
\bibitem[Reggiani et al.(2014)]{Reggiani14} Reggiani, M., Quanz, S.~P., Meyer, M.~R., et al.\ 2014, ApJL, 792, L23 
\bibitem[Richard et al.(2016)]{Richard16} Richard, S., Nelson, R.~P., \& Umurhan, O.~M.\ 2016, \mnras, 456, 3571 
\bibitem[Rosotti et al.(2016)]{Rosotti16} Rosotti, G.~P., Juhasz, A., Booth, R.~A., \& Clarke, C.~J.\ 2016, \mnras, 459, 2790 
\bibitem[Sallum et al.(2015)]{Sallum15} Sallum, S., Follette, K.~B., Eisner, J.~A., et al.\ 2015, Nature, 527, 342 
\bibitem[Shabram \& Boley(2013)]{SB13} Shabram, M., \& Boley, A.~C.\ 2013, ApJ, 767, 63 
\bibitem[Suzuki et al.(2010)]{Suzuki10} Suzuki, T.~K., Muto, T., \& Inutsuka, S.-i.\ 2010, \apj, 718, 1289 
\bibitem[Suzuki \& Inutsuka(2014)]{Suzuki14} Suzuki, T.~K., \& Inutsuka, S.-i.\ 2014, \apj, 784, 121 
\bibitem[Szul{\'a}gyi et al.(2014)]{Szulagyi14} Szul{\'a}gyi, J., Morbidelli, A., Crida, A., \& Masset, F.\ 2014, ApJ, 782, 65
\bibitem[Szul{\'a}gyi(2015)]{Szulagyi15} Szul{\'a}gyi, J., PhD thesis: ``Gas accretion onto Jupiter-mass Planets", 2015, https://people.phys.ethz.ch/$\sim$judits/thesisszulagyi.pdf
\bibitem[Szul{\'a}gyi et al.(2016a)]{Szulagyi16} Szul{\'a}gyi, J., Masset, F., Lega, E., et al.\ 2016, MNRAS, 460, 2853
\bibitem[Szul{\'a}gyi \& Mordasini (2016)]{SzM16} Szul{\'a}gyi, J. \& Mordasini, C., 2016, MNRASL,  arXiv:1609.08652
\bibitem[Szul{\'a}gyi et al.(2016b)]{SzMayer16} Szul{\'a}gyi, J. \&  Mayer, L., 2016, \mnras, arXiv:1610.01791
\bibitem[Tanaka et al.(2002)]{Tanaka02} Tanaka, H., Takeuchi, T., \& Ward, W.~R.\ 2002, \apj, 565, 1257 
\bibitem[Tanigawa et al.(2012)]{Tanigawa12} Tanigawa, T., Ohtsuki, K., \& Machida, M.~N., 2012, ApJ, 747, 47 
\bibitem[Tanigawa et al.(2014)]{Tanigawa14} Tanigawa, T., Maruta, A., \& Machida, M.~N.\ 2014, \apj, 784, 109 
\bibitem[Woitke et al.(2009)]{Woitke09} Woitke, P., Kamp, I., \& Thi, W.-F.\ 2009, \aap, 501, 383 
\bibitem[Zhu et al.(2013)]{Zhu13} Zhu, Z., Stone, J.~M., \& Rafikov, R.~R.\ 2013, \apj, 768, 143 
\bibitem[Zhu(2015)]{Zhu15} Zhu, Z.\ 2015, ApJ, 799, 16
\bibitem[Zhu et al.(2016)]{Zhu16} Zhu, Z., Ju, W., \& Stone, J.~M.\ 2016, arXiv:1609.09250 
\end{thebibliography}
\end{document}